\documentclass[preprint]{aastex}

\usepackage{graphicx}

\shorttitle{Metallicities of W~UMa-type binaries}
\shortauthors{Rucinski et al.}

\begin{document}

\title{Spectroscopic metallicity determinations for W~UMa-type 
binary stars\footnote{Based on observations obtained at the
David Dunlap Observatory, University of Toronto.}}

\author{Slavek M. Rucinski} 
\affil{Department of Astronomy and Astrophysics, 
University of Toronto,
50 St. George St., Toronto, Ontario, M5S 3H4, Canada}
\email{rucinski@astro.utoronto.ca}
\author{Theodor Pribulla, J\'{a}n Budaj}
\affil{Astronomical Institute, 
Slovak Academy of Sciences, 059~60, 
Tatransk\'a Lomnica, Slovak Republic}
\email{[pribulla,budaj]@ta3.sk}

\begin{abstract}
This study is the first attempt to determine the metallicities of 
W~UMa-type binary stars using spectroscopy.
We analyzed about 4,500 spectra collected at the David Dunlap 
Observatory (DDO). 
To circumvent problems caused by the extreme
spectral line broadening and blending and by the relatively 
low quality of the data, all spectra were subject 
to the same Broadening Function (BF) processing to determine 
the combined line strength in the spectral window 
centered on the Mg~I triplet between 5080 \AA\ and 5285 \AA.
All individual integrated BF's were subsequently orbital-phase 
averaged to derive a single line-strength indicator for each star.
The star sample was limited to 90 W~UMa-type (EW) binaries with 
the strict phase-constancy of colors and without spectral contamination
by spectroscopic companions. The best defined results were obtained
for a F-type sub-sample 
($0.32 < (B-V)_0 < 0.62$) of 52 binaries
for which integrated BF strengths could be interpolated 
in the model atmosphere predictions.
The logarithmic relative metallicities, $[M/H]$, for the F-type 
sub-sample indicate metal abundances roughly similar 
to the solar metallicity, but with a large scatter
which is partly due to combined random and systematic errors.
Because of the occurrence of a systematic color trend 
resulting from inherent limitations in our approach,
we were forced to set the absolute scale of metallicities to 
correspond to that derived from the $m_1$ index of the Str\"{o}mgren
$uvby$ photometry for 24 binaries of the F-type sub-sample. 
The trend-adjusted metallicities $[M/H]_1$ are distributed 
within $-0.65 < [M/H]_1 < +0.50$, with the spread reflecting
genuine metallicity differences between stars.
One half of the F-sub-sample binaries
have $[M/H]_1$ within $-0.37 < [M/H]_1 < +0.10$, 
a median of $-0.04$ and a mean of $-0.10$, with a tail 
towards low metallicities, and a possible bias against very high
metallicities. 
A parallel study of kinematic data, utilizing the most reliable and
recently obtained proper motion and radial velocity data for 78 
binaries of the full sample,
shows that the F-type sub-sample binaries (44 stars
with both velocities and metallicity determinations) 
have similar kinematic properties to solar neighborhood, 
thin-disk dwarfs with space velocity component dispersions:
$\sigma U = 33$ km~s$^{-1}$, $\sigma V = 23$ km~s$^{-1}$
and $\sigma W = 14$ km~s$^{-1}$. 
FU~Dra with a large spatial velocity, 
$V_{\rm tot} = 197$ km~s$^{-1}$ and $[M/H]_1 = -0.6 \pm 0.2$,
appears to be the only thick-disk object in the F-type sub-sample. 
The kinematic data indicate that the F-type EW binaries 
are typical, thin-disk population stars with ages about 3 -- 5.5 Gyr. 
The F-type binaries which appear to be older than the rest
tend to have systematically smaller mass-ratios than most 
of the EW binaries of the same period. 
\end{abstract}


\section{Introduction}
\label{intro}

This investigation addresses heavy-element abundances 
(hereinafter called ``metallicities'') of W~UMa-type 
eclipsing binary stars (hereinafter called EW systems). 
EW binary systems are defined as very close binary stars with
orbital periods shorter than about one day\footnote{The actual
upper limit of the orbital period may be 1.3 -- 1.5 days,
as suggested by statistics for the galactic field data
\citep{ruc1998a}.}, consisting of solar-type (late-A to early-K) 
stars of identical surface temperature in spite of 
very different component masses. The mass ratios are observed to be always
different from unity, down to 0.1 or even slightly less\footnote{We 
use the spectroscopic definition of the mass-ratio 
defined to be less than unity, 
irrespectively of the depths of the respective eclipses. This matter
is sometimes considered controversial for the ``W-subtype'' 
of the EW binaries for which less massive components appear to
have slightly higher surface brightness. 
We disregard the photometric definition as irrelevant in our case.}. 
The definition of EW binaries 
based on the equality of surface temperatures 
bypasses the issue of the best theoretical 
model applicable to W~UMa-type binaries. We also use the 
name ``contact binary''  because it is virtually 
impossible to equalize the surface temperatures of the disparate
components without some form of a strong physical contact. 

EW binaries are relatively common in the solar neighborhood
with a frequency of about one per 500 FGK dwarfs
\citep{ruc2006}. Their ages are presumed to be rather advanced, 
which is inferred mostly from their occurrence along the Main 
Sequences of progressively older open \citep{KR93,ruc1998b} 
and globular \citep{ruc2000} clusters. It is generally assumed that 
the EW binaries form from moderately close binaries 
through angular-momentum-loss processes which
take from a few hundred million to a few billion years. They
are absent among young stars, with the youngest known
in the open clusters Be~33 and Preasepe with estimated
ages of 700 and 900 Myrs \citep{ruc1998b}. 
It is not clear if the main evolution channel is 
through evolutionary expansion of components in very close, but
detached binaries, and to what extent this channel is augmented by
shrinkage of the orbits due to the magnetic-wind induced
angular momentum loss. Also, it is unknown
if the mass-ratio is reversed when the mass-exchange interaction 
takes place, although several arguments strongly point in 
this direction \citep{step2006,step2009}. 
In dense clusters, triple-body interactions may give an additional 
channel for the formation of EW binaries. We set all these problems and
interpretations aside and concentrate on spectral observations 
to derive metallicities irrespectively of evolutionary mechanisms. 
We supplement these results by kinematic data which shed
light on the most likely ages. 

Stellar metallicities $[M/H]$ are conventionally expressed as 
the fractional abundance of metals relative to hydrogen and 
expressed  in logarithmic units relative to the Sun,
$[M/H]=\log (M/H) - \log (M/H)_{\odot}$. 
EW binaries consist of solar-type stars with metallicities expected
to be similar to those found among stars in the solar neighborhood.
Any substantial chemical changes induced by evolution
would be expected only for primaries of the most massive among
EW progenitors and only for the CNO elements (to which
our spectral window is blind); for the majority
of the progenitors, the p-p cycle is not expected to modify
the metal content. The only possibility of unusual chemical 
composition would be then through contamination by a highly 
processed material from a degenerate, currently invisible, but 
originally much more massive third star in the system. Thus,
the derived metallicities are expected to reflect the properties 
of the EW progenitors even if mass-transfer takes place during 
their evolution. 
For galactic stars metallicities are (weakly) correlated with ages 
\citep{Reid2007,FB2009} so that the results may shed light 
on the W~UMa progenitor ages. 

Kinematical properties of the solar-neighborhood 
stars are known to correlate with their age. 
The extensive study of the EW binaries by \citet{bilir2005} was
based on a sample of 129 objects with parallaxes
and proper motions taken mostly from the Hipparcos satellite
mission \citep{esa1997} and supplemented by center-of-mass
radial velocities from the first nine papers of the DDO series 
(see Section~\ref{ddo}). 
This sample -- which was much more heterogeneous than that presented
here through the inclusion of various other short-period binaries 
-- suggests ages between 
a kinematically young (500 Myr) moving group to the age of 
the old galactic field population of about 5.5 Gyr.
We re-address this matter by limiting the analysis to
EW binaries of our spectroscopic sample and by using the center-of-mass
velocities derived in the whole DDO survey, supplemented by
new, improved reductions of the Hipparcos satellite proper motion data. 

Section~\ref{ddo} describes the spectroscopic observations of the 
DDO program while Section~\ref {BFs} discusses the use of the 
Broadening Functions approach to derive integrated spectral line strengths.
In Section~\ref{efftemp} the strongest dependence of the strengths
on the effective temperature is discussed. Determinations of metallicities
$[M/H]$ are given in Section~\ref{determine}; 
associated uncertainties in the determination process are discussed in 
Appendices~\ref{appendix-A} -- \ref{appendix-C}. 
Section~\ref{Strom} describes 
independent determinations utilizing the $m_1$ index of the 
Str\"{o}mgren $uvby$ photometry while Section~\ref{results}
contains the results of an attempt to combine 
both methods of metallicity determinations. 
Section~\ref{velocities} discusses
the space velocities of the binaries discussed in this paper.
Section~\ref{evolution} places the results in the
context of the evolution of EW binaries while 
Section~\ref{conclusions} gives a summary of the paper and 
recommendations for further research.

\placetable{tab_papers}                  

\section{The star sample and observations}
\label{ddo}

\subsection{Definition of the sample}

The main data source for the present investigation is 
the spectroscopic radial velocity (RV) survey of close binaries 
conducted during the last decade of the David Dunlap Observatory's 
(DDO) existence, in the years 1998 -- 2008. 
This survey consists of 15 main papers of the DDO RV analyses series.
We give the list of all publications containing spectroscopic data 
for 163 binaries of the DDO survey in Table~\ref{tab_papers}.
In addition to the variable-star name, 
we give the ``publication identification number''
in the form ($n$:$m$), where $n$ gives the DDO paper number
and $m$ is the star number in that paper. 

All stars identified as W~UMa-type or contact 
binaries in the DDO papers have been considered in the 
selection for metallicity determination. 
Three groups of EW binaries have been excluded from the current 
sample and are marked in Table~\ref{tab_papers}:
\begin{description}
  \item[All objects with components of dissimilar temperature,]
as indicated by light curves known to show unequally deep eclipses.
This investigation does not include objects with any indications of
being semi-detached or with poor thermal contact.
While this limitation was mainly related to the orbital-phase
invariance of integrated broadening functions 
(to limit the continuum-normalization issues),
it resulted in a more homogeneous and better defined sample.
Besides, for binaries with components of unequal temperature, our
technique of Broadening Functions would not provide consistent 
line-strength measurements. 
The excluded binary systems are marked in italics
in Table~\ref{tab_papers}. 
   \item[Spectroscopically triple systems.] 
The continuum normalization may be affected 
by the presence of a sharp-line companion
as then its continuum and the pseudo-continuum of the 
binary do not add to unity in proportion to their
luminosities; this prevents evaluation of correct
respective line strengths. 
This effect was described in \citet{ruc2010a},
in the caption to Fig.\ref{fig_logS_err} of this paper; we discuss 
it in full in Appendix~\ref{continuum}. We excluded all EW binaries with 
third components contributing more than 
$\beta = L_3/L_{12} > 0.05$ to the observed spectra.
The names of the rejected 
binaries are underlined in Table~\ref{tab_papers}. Three
binaries for which $\beta \le 0.05$: 
HX~UMa (8:8), SW~Lac (10:5) and AG~Vir (11:10); are marked
by the slanted-cross (X) symbols in the figures of this paper.
The high frequency of occurrence of companions to EW binaries is
discussed in \citet{triples1}, \citet{triples2}, and \citet{triples3}.
   \item[Binaries with poor data or lost original spectra.] 
Two systems, V404~Peg (16:1) and HH~Boo (16:3)
had poor DDO data because of their faintness. 
For four binaries, LS~Del (1:3), EF~Dra (1:4), 
AP~Leo (1:7), and UV~Lyn (1:8) observed at the Dominion Astrophysical Observatory, 
Victoria, B.C., we have been unable to acquire the original spectra for re-processing. 
The excluded binaries are marked by square brackets around their names 
in Table~\ref{tab_papers}.
\end{description}

The DDO program included few detached and semi-detached
binaries so the first of the above rejection criteria was applied
automatically. However, to be able to say something 
about the detached binaries which do show the same temperatures 
at all phases, similarly as in the EW
binaries, we included in this study 
two detached binaries, V753~Mon (3:9) and V1130~Tau (8:6).
For stars close to the Main Sequence similarity of temperatures implies 
similarity of the masses. In this case, the mass ratios are, 
$q = 0.97$ and $q=0.92$, respectively. These two binaries
are marked in all figures by diamond symbols. 

After the exclusions described above, the sample analyzed in this paper
consists of 90 EW systems. 
Most of them (62 systems) are brighter than $V_{\rm max} = 10$.
The binaries which are fainter than $V_{\rm max} \simeq 9.5$
frequently have poor standard photometry data and unreliable 
color-index information, a circumstance which had a potential impact on
our analysis of metallicities (Section~\ref{efftemp}).

Practically all EW binaries analyzed in this survey fall into
the spectral types interval between late-A to early-K types,
i.e.\ with effective temperatures not higher than about 8,500~K 
and not lower than about 4,700~K. 
The luminosity classes had been 
found from the blue region classification DDO spectra to 
correspond roughly to dwarfs (class V), 
although the luminosity class was usually very hard to evaluate 
from our spectra
because of the complex appearance of broadened and blended lines. 
For the contact-binary geometry, the orbital period 
can be taken as a proxy of the component size; 
a spread of the orbital periods for a given temperature
by less than about 50\% to 70\% indicates 
that some components may be moderately evolved, but generally still 
within the width of the Main Sequence \citep{RD1997,ruc1998b}.

\subsection{The spectroscopic material}
\label{spectra}

The spectral window used during the DDO radial-velocity 
program was centered on the 5184 \AA\ line of the 
neutral magnesium triplet ``b'' at 5167.32, 5172.68, and 5183.60 \AA. 
The triplet region was found to be 
very convenient for radial velocity (RV) measurements 
for most EW systems observed at DDO because of its
sufficient strength for all spectral types 
later than about A2V -- A4V 
down to the latest spectral-type EW binaries that we observed.
For the F2V template spectrum that we used,
the strongest line in the spectral window 
was actually not a Mg~I triplet line, but
the line Fe~II at 5169 \AA\ (see Table~\ref{tab_lines} and 
Section~\ref{synopsis}). In terms of the summed strength
of the absorption lines in the spectral window in the F2V spectrum, 
this line contributes about 8\%, while all three lines of the Mg~I 
triplet contribute about 7\%.
These proportions rapidly change with the spectral type
as the magnesium triplet becomes stronger relative to iron lines
for lower temperatures.

\placetable{tab_lines}                

The width of the actual spectral window that was used 
during the DDO spectral survey changed 
depending on the combination of the available
CCD detector and the diffraction grating. 
The spectral resolving power of the spectra varied 
within $R = 10,000 - 17,000$, depending on the (nightly) selectable 
slit width and the grating. Typical $S/N$ values were at a level of 
10 to 50. For the current investigation, we
utilized the one-dimensional, rectified spectra
of the radial velocity program. However, for the sake of uniformity of the
material, we undertook a major re-determination of all, 
over 4,500 Broadening Functions for the same spectral window 
(5080 -- 5285 \AA) and at the same resolution of 11.8 km/s/pixel. 
During this exercise, we addressed the quality of the spectra
and rejected spectra which had been given low weight in the RV program.
The total material consisted of 4531 spectra for 90 binaries; 
after all rejections, this was reduced to a total of 4333 spectra, hence on 
average 48 spectra per star.

\section{BF integrals as metallicity indicators}
\label{BFs}

\subsection{The Broadening Functions technique}

Stellar metallicities are usually determined using high signal-to-noise 
and high resolution spectra. 
With rotational velocities of individual components typically 
$V \sin i > 100 - 150$ km~s$^{-1}$, and with blended profiles of both 
components frequently exceeding 600 km~s$^{-1}$, the EW binaries are
very difficult targets for spectral studies. 
There is no hope of contemplating any traditional approach 
in determination of metallicities and ever observing individual lines
in the optical-region spectra of EW binaries.
Reliable techniques of radial velocity (RV) determinations based on 
cross-correlation or linear deconvolution have been developed only
within the last few decades. However, 
RV determinations are much less demanding on the quality of spectra 
than studies of chemical content because they are
based on simultaneous radial-velocity {\it shifts\/} (but not strengths) 
of many spectral lines. All papers of the DDO series 
used the Broadening Function (BF) approach \citep{ddo7}. The BF approach 
is superior to the cross-correlation because it is a linear 
de-convolution process and thus gives directly interpretable 
results without any need for additional calibration. 

Here we use the BF strength somewhat similarly to the concept 
of the equivalent width:
The integral of the BF function represents the combined strengths 
for all lines within the spectral window.   
The BF integrals were averaged in the temporal (orbital phase) domain.
Thus, we utilize integration and averaging to extract
the information on the line strength from the low S/N, heavily broadened 
and blended spectra.

The Broadening Functions deconvolution process was described in 
\citet{ruc1992,ddo7,ruc2012} and then
extensively used in all papers of the DDO series.
It is very similar to the Least Squares Deconvolution (LSD)
method introduced by \citet{DC1997} and \citet{Donati1997},
but utilizes a stellar spectrum template rather 
than a model spectrum. This choice was made
(1)~to directly relate radial velocities of the DDO survey to the RV
standards; (2)~to have template and program spectra affected by
the same instrumental effects; and (3)~to include all spectral
lines, even the weakest ones, which could be missed in model
spectra.

\begin{figure}[ht]
\begin{center}
\includegraphics[width=11.5cm]{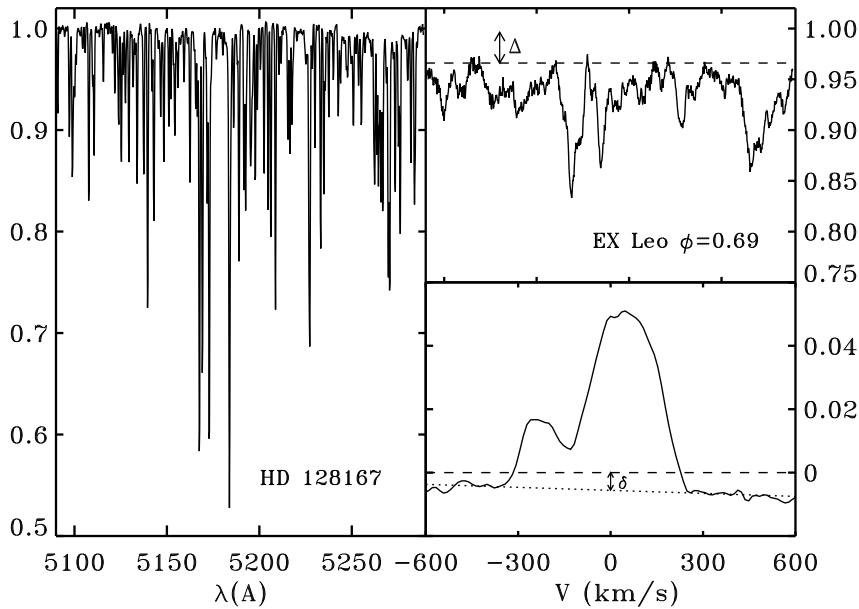}    
\caption{
\small 
Typical template and program spectra and the resulting BF.
The left panel shows the template spectrum (HD~128167), while
the right panel shows one of program spectra, here of
EX~Leo (4:5), at the orbital phase 0.69. 
The corresponding BF is plotted below. 
Note the negative shift in the BF baseline ($\delta$)
which is the result of spectral rectification to the pseudo-continuum
rather than to the real continuum. 
The integral of the BF above the baseline, $S$, is 
the quantity used to derive metallicities.
}
\label{fig_BF}
\end{center}
\end{figure}

\subsection{Synopsis of the metallicity determination}
\label{synopsis}

Derivation of the BF integrals, hereinafter called $S$, and metallicity
involves several steps and assumptions: 
\begin{description}
  \item[The spectral template.] From among about a dozen of 
template stars used during the DDO radial velocity survey, 
we selected the frequently observed 
HD~128167 ($\sigma$ Boo, HIP~71284, F2V). Its rotational velocity is
well below the resolution of our spectrograph of about 20 km~s$^{-1}$,
with measured values of $V \sin i$: 
9.3 km~s$^{-1}$   \citep{Valenti2005},
7.7 km~s$^{-1}$   \citep{Schro2009},
and 7.32 km~s$^{-1}$ \citep{Mart2010}.
HD~128167 permitted the best coverage of the whole range of 
the observed spectral types and $B-V$ colors in the sense that the
values of $S$ were rather uniformly distributed around unity.
When the spectral types and metallicities of 
a program object and of the template match perfectly, 
the $S$ integral should be equal to unity; a mismatch in the effective
temperatures has the strongest effect on $S$. 
Note that the template star actually
appears to show some metal under-abundance with $[M/H] = -0.38$
\citep{Valenti2005}, although this is unimportant 
because the template spectrum was used only as an intermediate
device in the BF determination process. 
The template spectrum obtained by averaging of several
individual spectra was very well defined and due 
to star's low rotational velocity, we did not encounter 
any problems with the continuum rectification as with 
the binary targets. 
  \item[Measurements of the strength.] 
Each rectified spectrum of the DDO program was de-convolved 
with the template spectrum producing one BF 
function which was subsequently integrated above
the local baseline to provide one value of $S$. As is shown in 
an example in Figure~\ref{fig_BF}, the BF baseline was 
typically shifted vertically by the (usually slightly negative)
value $\delta$. Such shifts represent missing information on the
real continuum level; in the
spectral domain the corresponding shifts are denoted $\Delta$ below. 
In an idealized case of the perfect spectral match of the program
and template spectra, and thanks to
the linear properties of the BF's, the integral of the BF above
the baseline should correctly represent the combined strength of the lines
within the spectral window. Then, a rectification step
involving a shift by $\Delta$ would produce
{\it linear\/} scaling of $S$ by $(1+\Delta)$, where 
$\Delta$ should be determinable from $\delta$. 
As we describe in Appendix~\ref{appendix-A},
we were not successful in our attempts to use the values of
$\delta$ to derive $\Delta$ in individual cases, but we could still use 
{\it the observed range in $\delta$ to estimate the 
expected range of systematic errors\/} in the measurements of $S$. 
  \item[Averaging of the BF strengths in phase.] 
This step allowed us to check on the constancy of the line strengths 
in phase, to ascertain the invariability of the apparent $T_{\rm eff}$ 
and thus to confirm the EW type of the binary.
Subsequently, individual values of $S$ were averaged with
a $2.5\,\sigma$ clipping which led to rejection of 4\% of all
spectra. None of the binaries of the final sample of 90 
binaries showed any phase dependence in $S$. 
   \item[Effective temperatures and colors.] The effective
temperature $T_{\rm eff}$ plays a crucial role because $S$
is very strongly dependent on it. Additionally, for 
contact binaries, many stellar properties correlate with
$T_{\rm eff}$. For practical reasons, we used the 
de-reddened color index $(B-V)_0$ as a proxy for $T_{\rm eff}$.
Derivation of the $B-V$ colors and interstellar reddening 
corrections is described in Section~\ref{efftemp}.
  \item[Calibration using model spectra.] 
The strengths of spectral lines as well as $S$ values 
depend on the atmospheric properties of the star, i.e.\ the effective 
temperature, $T_{\rm eff}$, the surface gravity, 
$\log g$, and the metallicity, $[M/H]$
($S = S(T_{\rm eff}, \log g, [M/H])$).
The dependence is strongest on $T_{\rm eff}$ and is much
weaker on $\log g$ and on $[M/H]$. 
While the model values of $S$ were found for two values of gravity,
$\log g = 3.5$ and 4.5 (cgs), the weak dependence on $g$ 
led us to use $S$ only for $\log g = 4.5$, but with the proper
account in the final error budget (see Section~\ref{models}
and Appendix~\ref{gravity}). 
For comparisons with the observations, the $T_{\rm eff}$ 
dependence was transformed into the $B-V$ color index 
dependence, as discussed in Section~\ref{efftemp}.
Note that the same template spectrum was used for
both the observed and model spectra, so that the template
served only in the process of converting the spectra into 
the Broadening Functions with no direct influence of its
properties on the final results.
  \item[Derivation of metallicities.] The $[M/H]$ values were
found by interpolation in the pre-computed table of the 
BF integrals $\log S = \log S(T_{\rm eff}, [M/H])$ for 
one value of $\log g = 4.5$, but taking into account 
the range of gravities in the error budget. 
We used two-dimensional interpolation utilizing the observed, 
reddening corrected values $(B-V)_0$ and the measured, 
average values of $\log S$.
We used $\log S$ instead of the combined line strength $S$ in 
anticipation of a need for scaling which would show up
as an additive shift rather than a multiplying factor. 
The curve of growth (dependence of the equivalent width on 
the abundance) has almost linear regions in logarithmic units which 
is an advantage when interpolation is used.   
Uncertainties and the error budget of metallicity 
determinations are discussed in detail in Appendix~\ref{appendix-A}.
\end{description}

\section{Effective temperature}
\label{efftemp}

\subsection{$\mathbf{B}-\mathbf{V}$ index data}
\label{bv}

The effective temperature of a star is the most important independent 
parameter in the spectral line-strength analysis. 
For EW binaries, it is the parameter known to correlate
with most of their astrophysical properties like for objects
distributed along the Main Sequence. 
The most obvious and simplest indicator of the effective temperature
would be the spectral type. Unfortunately, spectral
classification -- which we attempted to provide in most
of the DDO-series papers using the blue-region, low-resolution 
spectra -- is a very unreliable tool when applied to EW binaries. 
The large widths of spectral lines 
and the general washed-out appearance
of spectra compared with FGK-type spectral standards make
spectral typing uncertain. At best,
the spectral type can be estimated to two or three 
spectral sub-types. Systematic errors in spectral classification
for EW binaries are entirely unknown. 

The situation is not much better with color index data, mainly because 
literature data are very inhomogeneous or sometimes simply lacking.
Unfortunately, many light curves for EW's were obtained in 
instrumental systems without any attempts at standard-system calibration. 
There have been few studies aimed at providing 
calibrated $B-V$ or $b-y$ indices from observations at random
orbital phases. Even a few 
such observations are useful because the orbital color variations
for genuine W~UMa-type systems are small, typically by
less than 0.02 -- 0.03 in $B-V$. This circumstance simplifies 
the derivation of mean color indices although it prevents reduction 
of errors of such mean colors to less than about 0.01 -- 0.02. 

The $B-V$ color index was used as the temperature indicator 
in this investigation.
The main source of uniformly calibrated mean $B-V$ indices was
\citet{Terr2012} who obtained random $BVRI$ measurements specifically 
to improve the homogeneity of the EW binary photometric data. 
Their survey includes binaries fainter than $V \simeq 9$ of which 
42 are among the objects analyzed in this paper. We adopted the 
Tycho-2 \citep{esa1997} colors for the remaining binaries 
of the sample. The colors derived from the Tycho-2 project have
the advantage of uniformity in calibration over data 
gleaned from literature. The colors were obtained from the mean 
instrumental Tycho-2 magnitudes $B_T$ and $V_T$, transformed   
to the colors: $B-V = 0.85 \,(B_T-V_T)$. 
The uncertainties of the $B-V$ colors are much smaller than 
formal errors of mean magnitudes would imply because the latter 
include stellar variability; they are typically 
at a level of about 0.01 mag, but deteriorate 
rapidly beyond a $B_T$ of around 9.5 mag. 
The adopted values of $B-V$ are listed in Table~\ref{tab_main1}.

\placetable{tab_main1}                

The $B-V$ index can be
estimated from $b-y$ observations of EW binaries available in the 
literature. The EW binaries were observed in the Str\"{o}mgren 
photometric system by \citet{hildhill1975}, \citet{ruckal1981} and
\citet{ruc1983}. 
Additional studies which included EW binaries within 
larger photometric programs are cited in Section~\ref{Strom}.
The $uvby$ data have the advantage over the 
literature $B-V$ values of superior accuracy 
thanks to the well-defined and more rigorous 
calibration procedures. While we did not use the $b-y$ colors
directly, their agreement with the adopted $B-V$ data was
used for evaluation of the uncertainties of $B-V$. 

Interpretation of our spectroscopic data on metallicities
required reddening-corrected $(B-V)_0$ colors; the reddening corrections
are described in the next Section~\ref{reddening}.
The errors of the $(B-V)_0$ colors have been
estimated as resulting from three sources:
(1)~the scatter in the measured $B-V$; (2)~inconsistencies with 
spectral types estimated independently during the DDO 
program; (3)~uncertainties in the reddening corrections. 
They are listed in Table~\ref{tab_main1} with the observed 
mean $B-V$ and the reddening corrections $E(B-V)$. 

An attempt was made to use an infra-red color index in 
addition to $B-V$. The infrared observations are expected to give
better estimates of effective temperatures than $B-V$ because
they are less affected by interstellar
reddening. The {\it 2MASS\/} project \citep{2mass} provided
single-epoch, simultaneous observations in the $J$, $H$ and $K$ bands. 
The longest-base index $J-K$ can serve as an useful
effective temperature indicator for late-type stars; it is
metallicity independent and is weakly dependent on interstellar 
reddening with $E(J-K) \simeq 0.5 \times E(B-V)$. 
We found that the {\it 2MASS\/} 
data are not sufficiently precise for our purpose: They have 
errors at a level of typically $0.03$, but with large scatter in
between stars. 
Generally, the reddening corrected values of $(J-K)_0$ agreed with 
$(B-V)_0$ values, but we also noted large discrepencies 
for some objects which could be mostly traced to individual
poor $J-K$ measurements. 

\subsection{Interstellar reddening}
\label{reddening}

Most of the EW binaries of this study 
lie in the distance range 15 - 500 pc; the resulting
reddening corrections $E(B-V)$ in $(B-V)_0 = (B-V) - E(B-V)$
are generally small. Because of the very clumpy nature of the 
interstellar matter in the galactic disk, 
accurate (better than $< 0.01$ mag) values of $E(B-V)$ are relatively 
difficult to estimate; they become relatively more predictable 
at large distances through averaging in many individual clouds.  

We attempted to account for the nonuniform distribution
of the dust clouds in the Galaxy in the following way:
rather than using a model with one number characterizing
$E(B-V)$ per unit of distance, we utilized as primary information 
the maximum reddening correction 
in a given galactic direction, as seen throughout the whole Galaxy.  
Such maximum values were then scaled down by the estimated 
distance to the object assuming a finite thickness of the 
galactic dust layer. The values of $E_{\rm max}(l,b)$, interpolated 
from the maps of \citet{Schlegel998}, were found 
using the IDL routines provided by these 
authors\footnote{http://astro.princeton.edu/$\sim$schlegel/dust/index.html}.
Such maximum reddening values typically ranged between 
0.02 at the galactic poles to large values of several magnitudes
in the galactic plane. The assumed values of $E(B-V)$ were then
derived by scaling $E_{\rm max}(l,b)$
by the length of the path to the binary within a dust slab
of semi-thickness of 150 pc:
$E(B-V) = E_{\rm max}(l,b) \times \min(1, d \, \sin(b)/150)$. 
The distances $d$ were 
estimated in three different ways and then inter-compared to 
provide best estimates (given in Table~\ref{tab_main1}) using:
(1)~new determinations of the Hipparcos parallaxes 
\citep{leeu2007}; 
(2)~the colour-independent calibration,
$M_V = -1.5 - 12 \, \log P$, which applies 
to short-period EW's with $P<0.56$~d \citep{ruc2006}; 
and (3)~the color and reddening-dependent calibration, 
$M_V = -4.44 \, \log P + 3.02 \, (B-V)_0 + 0.12$, 
requiring a single-step iteration in the reddening-corrected 
color \citep{RD1997}. 

The values of $E{B−V}$ determined in the above way have 
a (relatively small) median of 0.027. Additional checks on the amount of 
reddening were possible for objects within 
30 pc using interstellar polarization data 
\citep{leroy1999} and the gas column density for the 
NaI~D-line \citep{Lall2003} and O~VI \citep{Oeg2005}  
interstellar lines, but no discrepancies from the 
previously determined values larger than 0.01 were noted. 

The reddening corrections were adjusted for 
two binaries, V1191~Cyg (13:2) and OO~Aql (12:1), with the
latter so uncertain that OO~Aql has been excluded from the 
metallicity analysis (Section~\ref{metal})\footnote{For
V1191~Cyg, we used $E(B-V) = 0.12$, based on the comparison
of the observed color with the spectral type.  
OO~Aql (12:1) has been recognized as an unusual object 
with the red color being due to its low temperature
\citep{moch81,hriv1989}, but the DDO observations 
\citep{ddo12} gave too large masses for a K-type binary;
the spectral corresponding to the total mass is F8V.
We assumed $E(B-V)=0.16$, which is
much larger than the 0.07 predicted from the sky position and 
assumed distance. 
Reddening is very clumpy in this area of the sky; a value as 
large as $E(B-V) \simeq 0.20$ is not excluded.}. 
We list the adopted $B-V$
colors and reddening estimates in Table~\ref{tab_main1}.
We note that the reddening
corrections, even if small, affect not only
the $(B-V)_0$ colors, but also the Str\"{o}mgren photometry
indices $m_1$ which were used for comparison and calibration
of our metallicity determinations 
(Section~\ref{Strom}).

\section{Determination of metallicities}
\label{determine}

\subsection{Model spectra}
\label{models}

A grid of synthetic spectra covering the 5070 -- 5300 \AA\ 
was computed using the code Synspec 
\citep{Hub1995,Krt1998} with the structural variables of the
model atmospheres (depth dependences of the state variables) adopted from
\citet{Cast2004}. The models assume LTE, radiative and
hydrostatic equilibrium with convection, and a plane-parallel geometry.
The grid was defined by 11 temperature, 2 gravity and 6 metallicity
values, with one value of micro-turbulence of 2~km~s$^{-1}$.
The effective temperatures, $T_{\rm eff}$, varied from 4000~K to 9000~K
in steps of 500~K. The two values of the surface
gravity were $\log g = 3.5$ and $\log g = 4.5$ (cgs) while the
metallicity, $[M/H]$, varied
from $-2.0$ to +0.5 in steps of 0.5 (i.e.\ metal abundances ranging
from 0.01 solar to 3-times solar). The abundances of all elements
(except of H, He) were varied in the same proportion relative to
the hydrogen abundance.
The spectra were calculated with a wavelength step of
0.01 \AA\ or about 0.6~km~s$^{-1}$ for our spectral window.

The synthetic spectra were prepared for use 
with the observed BF strengths $S$ by the following 
transformations:
(1)~We convolved them with a Gaussian profile with 
$\sigma = 15$ km~s$^{-1}$ to avoid problems with model 
spectra having shaper lines than for the template; 
(2)~We sampled the resulting spectra to the same resolution as the
observational spectra (11.8 km~s$^{-1}$~pix$^{-1}$);
and (3)~We determined the Broadening Functions 
and their strengths for
the observational spectral window 5080 -- 5285 \AA\ 
using the same stellar spectrum template HD~128167. 

\placetable{tab_models}         

The tables of $\log S_{\rm th}(T_{\rm eff}, [M/H])$
were computed for the two model values of $\log g$, but
we used only those for $\log g = 4.5$ (cgs).
Table~\ref{tab_models} lists the values of $\log S_{\rm th}$ for 
the observational $B-V$ scale of Bessell,
as is more fully described in the next Section~\ref{colors}. 
The influence of the gravity difference between $\log g = 3.5$ and 4.5
is not negligible and has been taken into account by including
a term in the total error budget (see Appendix~\ref{gravity}). 
In principle, we could determine gravities for individual W~UMa binaries,
but this would require us to provide
combined photometric (light curve) and spectroscopic (radial
velocities) solutions which we tried to avoid.
While we do have new radial-velocity solutions for all binaries
of the survey, lack of photometric data would eliminate some systems and
the combined solutions would have to include new, photometric-model
dependent uncertainties.

\begin{figure}[ht]
\begin{center}
\includegraphics[width=9cm]{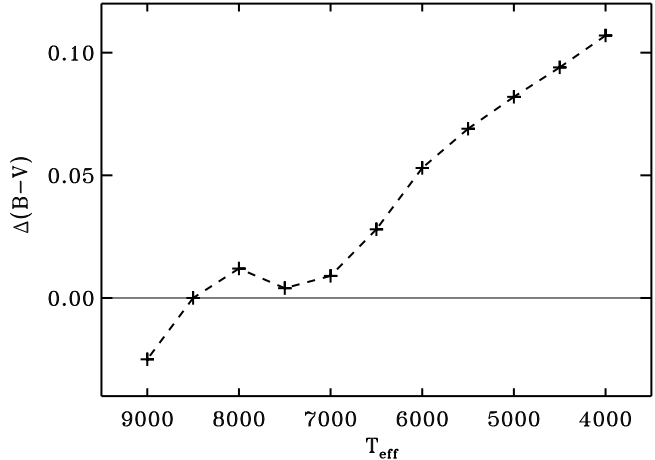}    
\caption{
\small 
Differences between $B-V$ values from the
\citet{bess1979} calibration and our
model results, plotted versus the model effective temperature for
the solar metallicity and $\log g = 4.5$. We  
used these color differences as one of indicators
of systematic uncertainties in the combined line strength estimates
($\epsilon_4$). 
}
\label{fig_B-V_dbv}
\end{center}
\end{figure}

\subsection{Synthetic $\mathbf{B}-\mathbf{V}$ colors}
\label{colors}

The same atmosphere models used for the computation
of the high resolution synthetic spectra for the Mg~I window
were used to calculate theoretical values of the $B-V$ color indices
using the filter transmission curves of the Johnson system
\citep{ubv1951}, and the zero-points for the magnitude system
as described in \citet{budaj2005}. 
The theoretical $B-V$ indices obtained from the model
atmospheres are metallicity dependent through the changes of opacity 
accompanying changes of $[M/H]$. This dependence is
important because it couples with the direct metallicity dependence 
of the integrated line
strength $S$ on the temperature. We took it into account by
interpolating the model values of $\log S$  for
each metallicity into the common, solar-metallicity $B-V$ scale. 

We compared our $T_{\rm eff}$, $B-V$
calibration with the observational calibration 
of \citet{bess1979} for the solar abundance. 
The differences in $\Delta (B-V)$, 
in the sense (Bessell -- present models), are shown in 
Figure~\ref{fig_B-V_dbv}. 
We applied the same differences to our model $B-V$ 
results for all metallicities assigning 
them to imperfections in filter transformations. 
They represent an inherent (systematic) 
uncertainty in the models and will reflect in uncertainties
of $[M/H]$ which we denote as $\epsilon_4$ (see Appendix~\ref{gravity}).

\subsection{Metallicities from $\mathbf{\log S}$}
\label{metal}

\begin{figure}[ht]
\begin{center}
\includegraphics[width=11.5cm]{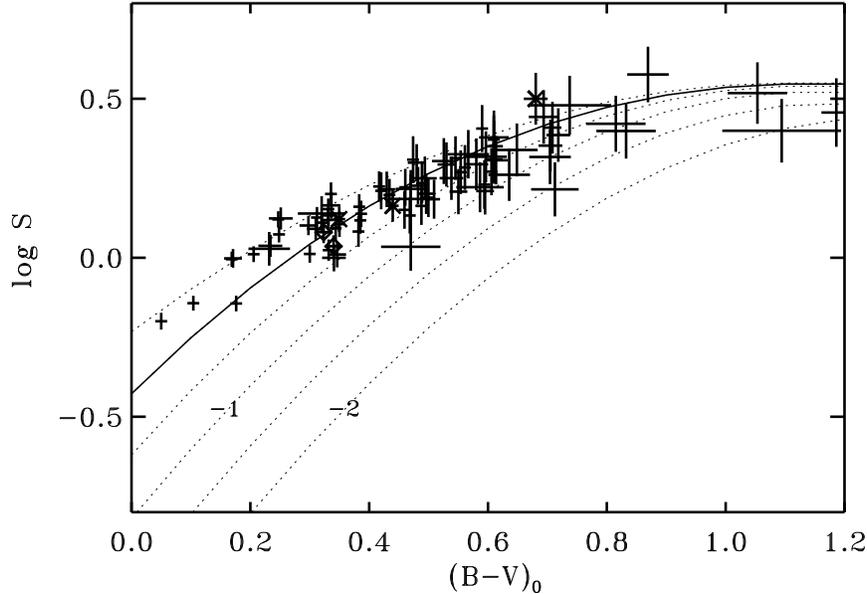}    
\caption{
\small 
The observed values of the BF integrals, $\log S$ (error bar symbols),
are over-plotted on the model results (lines)
for $\log g = 4.5$. The observational line strengths have been
corrected for the baseline displacements $\delta$, as described
in Appendix~\ref{continuum}. 
The dotted lines correspond to model results for
 metallicities $[M/H]$ in intervals
of 0.5 between $-2$ and $+0.5$; the solar case is shown by
the solid line. 
In this and the following figures, the two detached binaries 
are marked by the open rhombus symbols 
while the three triple systems with $L_3/L_{12}<0.05$ are marked
by X symbols. The error bars for $(B-V)_0$ and $\log S$ 
do not take into account uncertainties
in the model atmospheres and correspond to the
quadratically added $\epsilon_1$ and $\epsilon_2$ only. 
}
\label{fig_B-V_logS}
\end{center}
\end{figure}

Table~\ref{tab_main1} lists the observational data assumed 
for metallicity determination, Table~\ref{tab_models} gives
the model predictions and
Table~\ref{tab_main2} lists the integrated BF line strengths
$\log S$ of individual stars. 
Figure~\ref{fig_B-V_logS} shows the values of $\log S$
plotted on top of the model atmosphere results;
the data points with error bars represent 
the measured values while the model results are shown by dotted lines. 
The general impression is of a good agreement with
the solar-metallicity models.
The two detached binaries with identical 
components and the three triple systems with faint companions 
do not appear to deviate much from the general, solar-metallicity
dependence and seem to merge within the general scatter of the data. 

\placetable{tab_main2}         

Note that Figure~\ref{fig_B-V_logS}
does not present {\it metallicity determinations\/}; the
results only suggest a
general agreement with the assumption of the solar metallicity
with $[M/H]$ within roughly $\pm 0.5$. Determination of $[M/H]$ requires
a detailed comparison with the atmosphere models, with all
model uncertainties, and is more complex than what can be 
directly inferred from Figure~\ref{fig_B-V_logS}. 

The model predictions of the integrated line strengths $\log S$ 
(dotted lines in Figure~\ref{fig_B-V_logS}) form 
a smooth and simple two-dimensional function
$\log S = f((B-V)_0,[M/H])$.
As expected, the strengths $\log S$ depend monotonically 
on $[M/H]$ and on $B-V$. They tend to saturate for red
stars and for positive values of $[M/H]$.
This simple monotonic shape of the model function $f$
permits its inversion into a function $[M/H] = F((B-V)_0, \log S)$ to
use for determination of $[M/H]$. 
However, such an inversion is constrained by the tendency for saturation 
at red colors and large values of $[M/H]$. Besides, determination of $[M/H]$ 
requires that uncertainties in the model function are taken
into account, in addition to uncertainties in the observational
data. Such systematic errors, denoted as $\epsilon_3$ and $\epsilon_4$
and the related quasi-random error $\epsilon_5$, are 
discussed in Appendix~\ref{errors}.  

\begin{figure}[ht]
\begin{center}
\includegraphics[width=11.5cm]{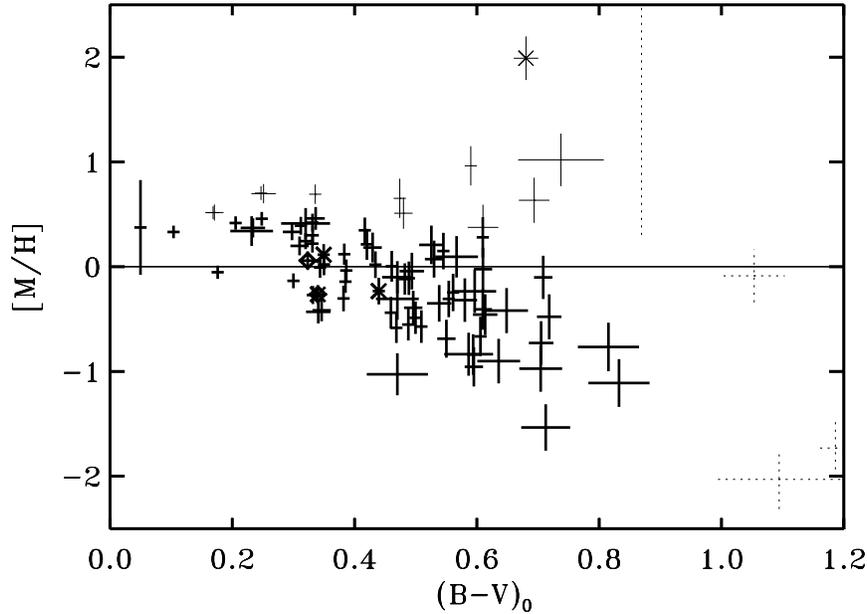}    
\caption{
\small 
Determinations of $[M/H]$ based on the measured BF strengths 
$\log S$. The horizontal error bars correspond to
$\epsilon (B-V)$ while the vertical error bars combine all
components of the $\log S$ error, as described in 
Appendix~\ref{errors}, added in quadrature.
The remaining symbols are as in Figure~\ref{fig_B-V_logS}.
Very large, positive values of $[M/H]$
mainly result from linear extrapolation beyond 
the model atmosphere grid beyond $[M/H] = +0.5$ 
(see Figure~\ref{fig_B-V_logS}); such extrapolated 
values are marked by thin symbols. 
Also uncertain are determinations for $(B-V)_0 > +0.85$
(thin dotted line symbols).
}
\label{fig_FeH}
\end{center}
\end{figure}

The individual values of $[M/H]$ obtained by bi-linear interpolation
in the model data are shown in Figure~\ref{fig_FeH}. 
The EW binaries appear to have metallicities very roughly solar,
but with systematic and significant deviations: 
\begin{itemize}
   \item Some binaries occupy the region in Figure~\ref{fig_FeH} 
of high metallicities. Practically without exception, they result from 
extrapolation beyond the largest model value of $[M/H] = +0.5$ 
exacerbated by large uncertainties in $\log S$, particularly at 
red colors. We were unable to determine metallicities for 
such binaries; this ``saturation'' effect happened for 12 
binaries.
   \item Because of the shape of the model function 
$[M/H] = F(B-V, \log S)$, metallicity determinations for
red binaries with $(B-V)_0 > +0.85$ are entirely unreliable. 
Four such binaries were eliminated.
   \item A trend of progressively lower metallicities
for later spectral types extends over the whole range of 
colors. It starts at positive values of $[M/H]$ 
for blue binaries and continues to negative metallicities for
red binaries. The trend is most likely caused by the problems 
with pseudo-continuum placement
discussed in Section~\ref{BFs} and summarized in 
Appendix~\ref{appendix-A}.
We do not see such a color trend in results based
on the Str\"{o}mgren photometry $m_1$ index 
for a sub-sample of F-type systems (Section~\ref{Strom}). 
   \item Early-type systems seem to show metallicities higher than
solar. This may be due to the same systematic trend as  
above or it may result from the assumption of low,
solar micro-turbulence value for our models: 
Lateral pressure gradients in the outer layers may generate
large turbulent velocities which may in turn lead to 
line strengths systematically larger than effective 
temperatures would imply.
\end{itemize}

In Table~\ref{tab_main2}, the values of $[M/H]$ for binaries
suffering from the ``saturation'' are given in square parentheses;
they are marked by thin symbols and error bars in the subsequent 
figures. Removal of such binaries may produce a bias in the 
final results against positive metallicities;
this bias is one of the deficiencies of our analysis. We discuss
determinations of $[M/H]$ in full in Section~\ref{results}
after a description of literature results on metallicities
derived from Str\"{o}mgren photometry.

\section{Str\"{o}mgren photometry: The literature data}
\label{Strom}

It is not our intention to present a study of  
Str\"{o}mgren photometry results because the literature data
are available for only 35 EW binaries of our sample
and because we feel that such a useful and relatively
easy to conduct survey should be done with our results as an incentive. 

The $m_1 = (v-b)-(b-y)$ index of the Str\"{o}mgren photometry
has previously been used to study the metallicity of W~UMa-type binaries;
we disregarded the index $c_1 = (u-v - (v-b)$ because of
its sensitivity to the gravity. 
\citet{ruckal1981} obtained a few $uvby$ observations per 
star and analyzed them together with similar fragmentary data from
\citet{hildhill1975}. They showed that the EW binaries 
appear to be Population~I objects. 
Some systems were found to show elevated continuum levels 
in the $u$ and $v$ bands which could be assigned to chromospheric 
activity. Chromospheric emission would modify the $m_1$ index
to mimic a decreased metal-line blanketing. A subsequent study
of southern EW's \citep{ruc1983}
showed that chromospheric emissions may be of concern 
only for the latest spectral types. Therefore, the 
UV excesses observed in some contact binaries are probably
caused by genuine differences in metallicity, as 
originally suggested by \citet{eggen1967}. The spread in metallicities 
derived from deviations in $m_1$ from the standard sequence (see below) 
within $-0.02 < m_1 < +0.08$ would correspond to 
metal abundances in the range $-0.6 < [M/H] < +0.4$. 

\begin{figure}[ht]
\begin{center}
\includegraphics[width=10.0cm]{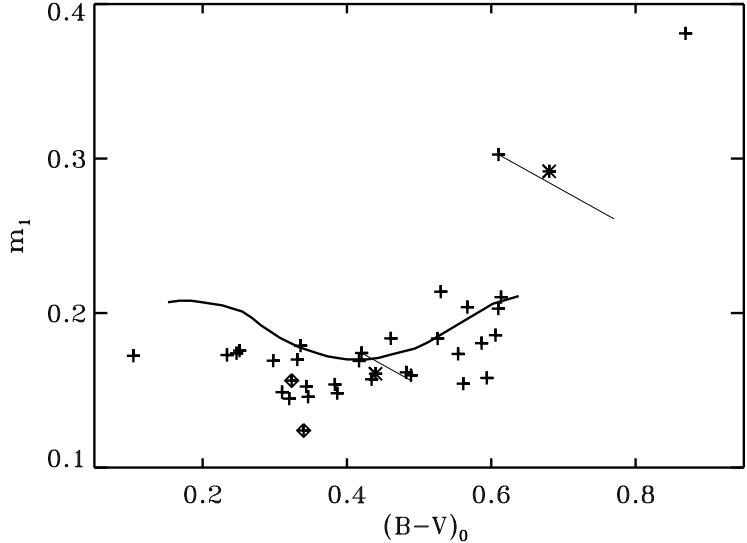}    
\caption{
\small 
The reddening-corrected Str\"{o}mgren indices 
$m_1$ from literature sources are shown by crosses 
versus the de-reddened color $(B-V)_0$. 
The triple and detached binaries included in this study are
marked by the same symbols as in previous figures. 
The reddening corrections are
shown by lines for the cases when $E(B-V)>0.05$. 
The line gives the Hyades standard relation 
for mid-A to early-G type dwarfs from \citet{Hyades1966}. 
}
\label{fig_m1}
\end{center}
\end{figure}

\placetable{tab_uvby}           

Additional Str\"{o}mgren photometry data for EW binaries could
be found in random observations obtained within large-scale 
surveys of brightest stars and stars in clusters by 
\citet{craw1970},
\citet{blaa1976},  
\citet{schm1976}, 
\citet{perr1982}, 
\citet{ols1983,ols1993,ols1994}, 
\citet{jord1996}.
So far, these data have not been analyzed in the EW metallicity context. 
The $uvby$ data collected from the literature 
are listed in Table~\ref{tab_uvby}. The observations were of different
quality and some entries in the table result from the averaging of
several observations. 

We estimated metallicity using the $m_1$ data corrected for interstellar
reddening: $m_1 = m_1 ({\rm obs}) + 0.26*E(B-V)$
\citep{stromgren1966}\footnote{The de-reddened index $m_1$ is
sometimes called $m_0$.}. The values of the de-reddened
$m_1$ are plotted in Figure~\ref{fig_m1} together with the standard
relation for the Hyades defined by \citet{Hyades1966}. 
The figure shows that most binaries have lower metallicities 
than that of the Hyades, $[M/H] = +0.13$  
\citep{Pins2004}.
Reddening corrections may in some cases influence the $m_1$
indices quite strongly, as in the cases of OO~Aql (12:1) and QX~And (15:1)
for which $E(B-V)$ is equal 0.16 and 0.07, respectively.
As metioned in Section~\ref{reddening}, the case of
OO~Aql is a special one so that we excluded
this binary from further considerations. 
Metallicity estimates $[M/H]_{\rm m}$ derived from 
the $m_1$ indices are described in Section~\ref{subF}.

\section{Results of metallicity determinations}
\label{results}

\subsection{The sub-sample of F-type binaries}
\label{subF}

Discussion of metallicity determinations is facilitated when 
the sample is divided into three sub-samples, 
with the middle one consisting of the dominant F-type systems.
There are 57 binaries in this sub-sample with 
spectral types within F1 -- G2 and colors within
$0.32 < B-V < 0.62$; 52 of these binaries have well determined
metallicities, i.e.\ not exceeding $[M/H] = +0.5$. 
Metallicities determined for the red and the 
blue sub-samples could not be related to the $m_1$-based results
and are not discussed in the following. 
The reason that as many as almost two thirds of the sample are
F-type EW binaries is a combination of the 
relatively shallow depth of the DDO survey and of
the shape of the luminosity function for the solar
neighborhood stars.

\begin{figure}[ht]
\begin{center}
\includegraphics[width=10.0cm]{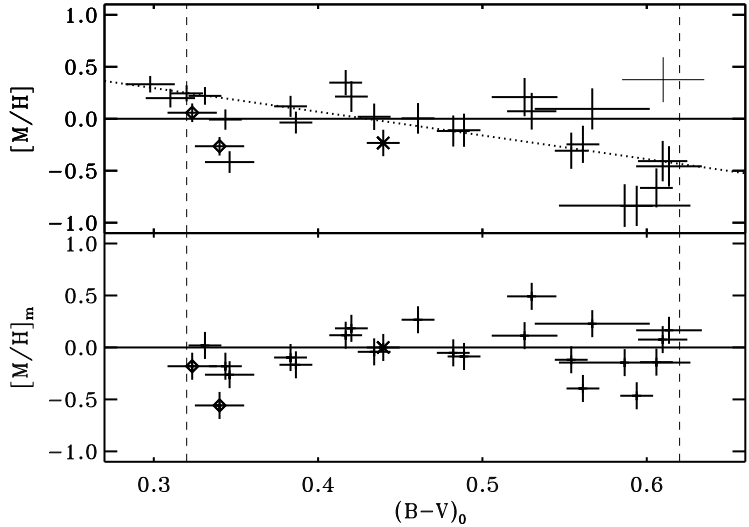}    
\caption{
\small 
Comparison of our spectroscopic metallicity results 
$[M/H]$ (upper panel) with metallicities $[M/H]_{\rm m}$ estimated from 
the $m_1$ index, as described in the text 
(lower panel). 
The dotted line in the upper panel gives the linear fit 
to the differences between the BF-based and $m_1$-based
determinations for 24 F-type binaries within 
$0.32 < (B-V)_0 < 0.62$. 
OO~Aql (12:1) at the right end was not 
used in the fit (see the text); it is marked by a thin symbol.
}
\label{fig_Fe_m1}
\end{center}
\end{figure}

Metallicities for the F-type sub-sample of the EW binaries
were derived using the calibration of  
\citet{craw1976},
$[M/H]_{\rm m} = [M/H]^{\rm Hyades} - 13 \times \delta m_1$, with
$\delta m_1 = m_1^{\rm Hyades} - m_1$,
using the updated metallicity of Hyades, 
$[M/H]^{\rm Hyades}=+0.13$ \citep{Pins2004}. 
This calibration cannot be used for stars with $(B-V)_0 < 0.32$  
because then a $c_1$-index dependent term should be included.
$c_1$ is particularly sensitive to
gravity, a dependence which we disregarded, and can be
more affected by ultraviolet chromospheric emission than $m_1$. 
Because of the inhomogeneity of
the literature $uvby$-data, we did not use any observational errors
and uniformly set the same error for the $m_1$-derived
metallicities, $\epsilon [M/H]_{\rm m} = 0.13$. This number
is approximate and assumes that the typical error of $m_1$ is 0.01, 
and results from the accumulation of observational, reddening-correction
and Hyades-sequence subtraction errors.
The comparison of the BF-based and $m_1$ based $[M/H]$
determinations is shown in Figure~\ref{fig_Fe_m1}
for 24 binaries which have determinations utilizing both methods.

\begin{figure}[ht]
\begin{center}
\includegraphics[width=10.0cm]{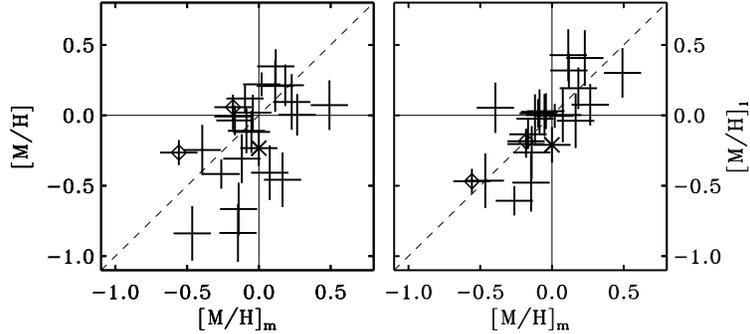}    
\caption{
\small 
The two panels show correlations of the BF-based metallicities 
with the $m_1$-based ($[M/H]_{\rm m}$) metallicities for the EW binaries
of the ``F-type sub-sample'' ($0.32 < (B-V)_0 < 0.62$). 
The vertical axes of both panels show $[M/H]$
and the trend-corrected $[M/H]_1$ (see Section~\ref{notrend}, 
respectively. 
}
\label{fig_Fe_m1_comp}
\end{center}
\end{figure}

Figure~\ref{fig_Fe_m1} shows a color trend in 
the BF-based values of $[M/H]$. 
This is the same trend as in Figure~\ref{fig_FeH} 
(Section~\ref{metal}), but within
a shorter color range. No such trend is obvious
in the $m_1$-based values $[M/H]_{\rm m}$. 
The values of $[M/H]$ and $[M/H]_{\rm m}$ correlate rather weakly, as
is visible in the left panel of Figure~\ref{fig_Fe_m1_comp}. 
The Pearson correlation coefficient 
is $C_P = 0.48 \pm 0.03$, while the Spearman rank coefficient
is $C_S = 0.53 \pm 0.01$. The positive correlation 
indicates that the deviations from zero metallicity are most
probably real, but the significance of the correlation is low. 
We stress that the two methods are entirely independent. 

\begin{figure}[ht]
\begin{center}
\includegraphics[width=11.5cm]{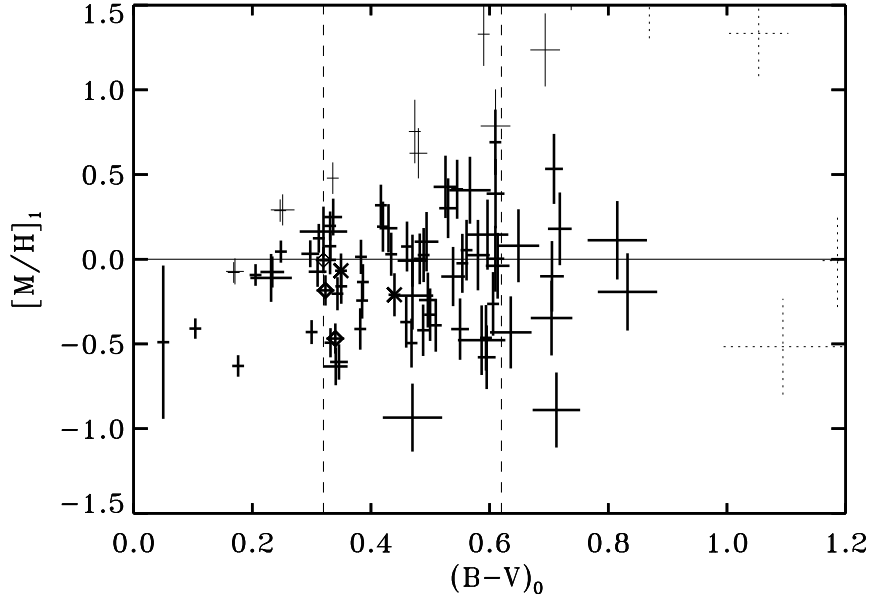}    
\caption{
\small 
Metallicities $[M/H]_1$ obtained from the 
BF-based values of $[M/H]$, after the
correction for the trend determined within the color range
$0.32 < (B-V)_0 < 0.62$,
as in Figure~\ref{fig_Fe_m1}. The values of $[M/H]_1$ 
are explicitly tied to the $m_1$-based system.
The thin symbols are for EW systems for which 
the BF line strengths could not provide reliable metallicity 
estimates; these data points are shown for illustration 
only and have not been used in our discussion of the EW metallicities. 
}
\label{fig_B-V_Fe3}
\end{center}
\end{figure}

\subsection{Correction for the color trend}
\label{notrend}

We decided to remove the color trend in the determinations
of $[M/H]$ by fitting and subtracting 
a linear color dependence based on 24 binaries 
of the F-type sub-sample which have reliable metallicities derived 
using spectroscopic and photometric methods. We had a choice here:
(1)~the fit could be fully independent of the $m_1$-derived 
results, or (2)~we could directly relate the BF-based results to the 
Str\"{o}mgren photometry results by fitting the individual differences. 
We decided to follow the latter
route and fitted a linear $(B-V)_0$ dependence to the differences, 
$[M/H]_{\rm trend} = [M/H] - [M/H]_{\rm m}$. Subtraction of such 
a relation brings the BF-derived data
into the system established by the $m_1$ index preserving 
information on metallicities of individual objects. 
The trend is given by:
$[M/H]_{\rm trend} = (-0.05 \pm 0.04) - (2.28 \pm 0.40) 
\times [(B-V)_0 - 0.45]$. 
The standard deviation for a single difference (after
allowing for two degrees of freedom) is 0.19 which gives 
independent information on the combined uncertainties of both
methods of the metallicity determination. To add up quadratically
to 0.19, with the assumed error $\epsilon [M/H]_{\rm m} = 0.13$,
the error of the BF-based determinations would have to be
$\epsilon [M/H]_1 \simeq 0.14$. This is consistent with  
the median (0.16) of individual errors derived 
through propagation of individual uncertainty contributions 
(see Appendix~\ref{errors} and Table~\ref{tab_main2}). 

The values of $[M/H]_1 = [M/H] - [M/H]_{\rm trend}$ show a tighter 
correlation with the $m_1$-based results than the uncorrected
metallicities $[M/H]$, with the respective correlation coefficients,
$C_P = 0.74 \pm 0.02$ and 
$C_S = 0.73 \pm 0.01$ (see the previous section). 
The correlation is
shown in the right panel of Figure~\ref{fig_Fe_m1_comp}.
We list the values of $[M/H]_1$ 
in Table~\ref{tab_main2} and show them versus $(B-V)_0$ in
Figure~\ref{fig_B-V_Fe3}. They may be considered
to be the final results of this analysis in the sense that they should
correctly reflect differences in individual metallicities 
among stars.
While we determined them for all binaries of this study
extending the trend to all colors,
they are strictly meaningful for the F-type sub-sample 
and for values $[M/H]_1$ which were obtained without extrapolation 
beyond the model grid with $[M/H] > +0.5$. 

It is useful to compare Figure~\ref{fig_B-V_Fe3} 
with Figure~\ref{fig_FeH}. The data points apparently follow
the horizontal line corresponding to the solar value. We stress
that, in terms of the zero-point, the values of
$[M/H]_1$ are entirely dependent on the $m_1$ calibration of 
\citet{craw1976} for the F-type sub-sample.
In addition, removal of the trend makes the results 
for the few early-type ($(B-V)_0 < 0.25$) binaries 
ambiguous: The metallicities as determined initially, i.e.\
$[M/H]$, were mostly positive, but 
after the trend has been removed, they are now negative in $[M/H]_1$. 
Possibly, the trend correction should not extend
into the domain of early-type EW binaries: The
continuum placement problems are practically absent for 
these stars while the integrated line strengths $\log S$ are
well defined and not very small, typically at 
about 1/2 strength of that for the template 
(see Figure~\ref{fig_B-V_logS}). 
We simply cannot state anything firm 
about metallicities of the early-type binaries. 
Even less trustworthy are metallicity determinations for
stars with spectral types later than G2. 

\begin{figure}[ht]
\begin{center}
\includegraphics[width=10.0cm]{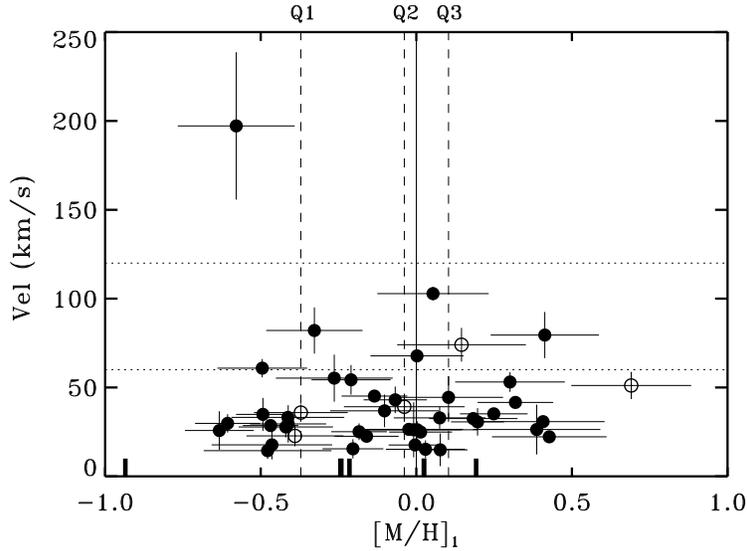}    
\caption{
\small 
The relation between the metallicity index $[M/H]_1$ and
the total spatial velocity for F-type stars (filled
circles), as described 
in Section~\ref{velocities}. Velocity data taken from 
\citet{bilir2005} are marked by open circles.
The high velocity binary FU~Dra is the single point in the
upper left of the figure. The 60 and 120 km~s$^{-1}$ lines
delineate (arbitrary) velocity ranges to distinguish stars
of small, moderate and large spatial velocities. 
Binaries without spatial velocities are marked by
thick ticks along the lower horizontal axis.
The quartiles of the distribution of $[M/H]_1$ for 52
F-type binaries are marked along the upper horizontal axis 
and shown by vertical broken lines.
}
\label{fig_vels_Fe}
\end{center}
\end{figure}

\subsection{Metallicities of the F-type sub-sample binaries}
\label{Fsample}

The correlation observed for the F-type sub-sample 
between the trend-corrected metallicities 
$[M/H]_1$ and the  $m_1$-based metallicities
$[M/H]_{\rm m}$ indicates that {\it both\/} 
techniques provide meaningful results. 
Thus, it strongly supports the view that Str\"{o}mgren photometry
$\delta m_1$ deviations are indeed due to metallicity 
differences and that they are not affected by
chromospheric activity. 
While the inter-comparison of the BF-based and $m_1$-based 
results and determination of the color trend utilized
24 binaries in common, there are more than twice as many F-type 
binaries (52) with well determined spectroscopic metallicities $[M/H]_1$.
This number is, however, still too small for a statistical analysis.
The distribution of the trend-corrected metallicities $[M/H]_1$ 
is shown in Figure~\ref{fig_vels_Fe}.

The $[M/H]_1$ distribution for all 52 F-type binaries extends 
within $-0.65 < [M/H]_1 < +0.5$, but the width is partly
affected by the large determination errors 
(the median $\epsilon [M/H] = 0.16$, see Section \ref{appendix-A}). 
To reduce their influence, the distribution 
has been characterized by low-order moments:   
The mean($[M/H]_1$) $= -0.10$, the dispersion $\sigma{[M/H]_1} = 0.33$,
and the mean absolute deviation $\delta{[M/H]_1} = 0.26$.
The distribution is asymmetric with a tail towards low metallicities:
The lower quartile is $Q_1([M/H]_1) = -0.37$,
the median $Q_2([M/H]_1) = -0.04$ and
the upper quartile $Q_3([M/H]_1) = +0.10$. If we disregard
the asymmetry and assume that the observational errors broaden
the $[M/H]_1$ distribution by 0.16 (the median of all 
individual, propagated errors), then -- formally --
the full dispersion of $[M/H]_1$ is reduced to 0.29.
Thus, in the simplest terms, the distribution can be characterized
as $\overline{[M/H]_1} = -0.1 \pm 0.3$.
Keeping in mind that our result is entirely dependent 
on the adjustment to the $m_1$-based metallicity scale,
the small value of the mean metallicity and their
relatively narrow observed range for the F-type 
sub-sample of the EW binaries is consistent with typical 
metallicities in the solar neighborhood where 
the age range from the present to 10 Gyr corresponds to 
metallicities ranging from +0.15 to $-0.3$ \citep{Reid2007}. 

The binaries with $[M/H]_1$ outside the $Q1 - Q3$ quartile range by 0.1,
that is outside $[-0.47,+0.20]$, are listed below. 
The pairs of numbers in the square brackets 
give $[M/H]_{\rm m}$ and $[M/H]_1$; their differences can be used 
to appreciate the level of uncertainties in the
metallicity determinations.
\begin{description}
\item[Large metallicity,] $ \mathbf{[M/H] > +0.20 }$:
   V839~Oph (2:8)  [+0.49, +0.30],
   V2377~Oph (4:8) [+0.11, +0.43], 
   V335~Peg (9:10) [+0.12, +0.32],
   W~UMa (12:9)    [+0.23, +0.41].
\item[Low metallicity, ] $ \mathbf{[M/H] < -0.47 }$:
   DK~Cyg (2:3)   [$-0.26$, $-0.61$],
   V1130~Tau (8:6)[$-0.56$, $-0.47$] (detached binary),
   AM~Leo (12:5)  [$-0.15$, $-0.48$].
\end{description}

As mentioned before, only about one half (24) of all F-type
binaries with reliable metallicity determinations (52) 
have Str\"{o}mgren photometry data. 
Among the binaries with only spectroscopic 
metallicities, we note several low-metallicity 
outliers with $[M/H]_1 < -0.47$: 
XZ~Leo (2:7), FU~Dra (3:5), GM~Dra (6:5), FP~Boo (10:4) 
and V1003~Her (13:2)
with $[M/H]_1 = -0.63$, $-0.58$, $-0.50$, $-0.49$ and $-0.93$. 
Among these stars, FU~Dra
is a high velocity star
($V_{\rm tot} = $ 197 km~s$^{-1}$), apparently the only such
one in the analyzed sample (Section~\ref{velocities}).

\section{Kinematic properties of EW binaries}
\label{velocities}

Space velocity is the only other independent 
observable quantity which depends on stellar age. In this section
we describe an attempt to correlate metallicities 
with the spatial velocities of the EW binaries. 
One expects that most of the 
binaries belong to the disk-population which
dominates in the solar neighborhood. The oldest, highest-velocity, 
lowest-metallicity, halo or Population II 
stars are very rare, with a relative 
frequency of only 0.1\% \citep{Weis1972,deJong2010}. Within the
disk population, most of the stars, about 93\% \citep{deJong2010}
to 96\% \citep{AB2009}, are {\it thin-disk\/} objects 
(scale-height $<300$ pc) with metallicities 
similar to the solar metallicity. According to \citet{Schon2009},
all thin-disk stars may be younger than 7~Gyr.
The {\it thick-disk\/} population stars \citep{gil1983}
appear in the solar neighborhood with a relative 
frequency of the remaining few percent; this number
is the subject of intensive research. They have distinctly
lower metallicities than the solar, typically 
$[M/H] \simeq -0.5$ to $-1.0$, but with a wide range extending 
down to metallicities as low as $-2.2$, i.e.\ including
objects formerly known as ``intermediate Population~II''
or ``47~Tuc population'' with ages  $>10$ Gyr. 
Velocities of disk population stars are well organized and reflect 
the global galactic rotation. Their dispersions increase 
over time \citep{Wielen1977} through dynamical
processes such as scattering by molecular clouds
and spiral arms and processes
following accretion of low-luminosity satellites by the Galaxy.
Thus, space velocities 
can be used statistically as an age indicator for a group of objects.
Allowance must be made for the spectral type dependence
because velocity dispersions increase for late-type stars
\citep{SG2007,AB2009,holm2009} as a result of longer survival
times for lower stellar masses.

\begin{figure}[ht]
\begin{center}
\includegraphics[width=11.5cm]{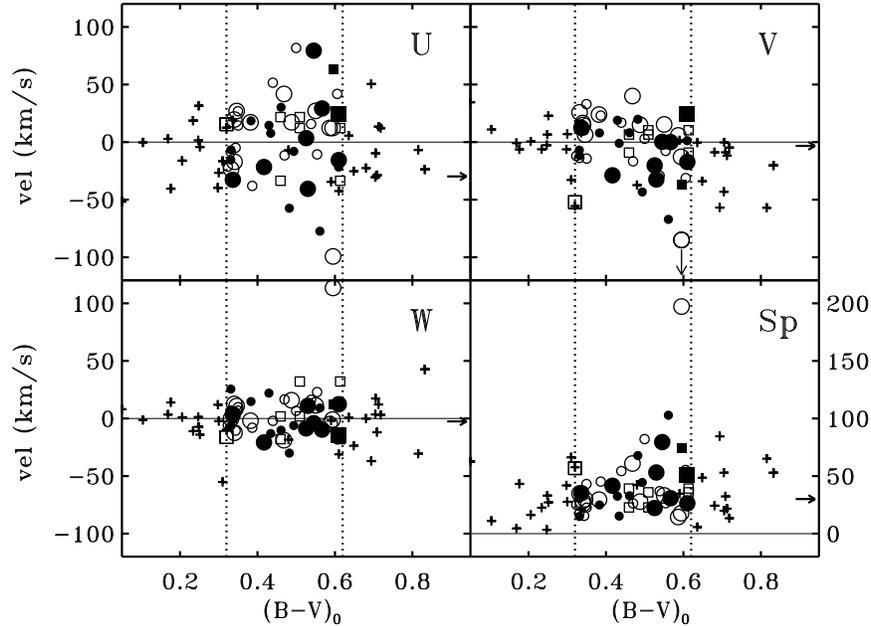}    
\caption{
\small 
The $UVW$ components of galactic velocities of the EW
binaries discussed in this paper (the three upper and left panels)
and the total spatial velocities (the lower, right panel).
Metallicities $[M/H]_1$ for the F-type binaries 
with $0.32 < (B-V)_0 < 0.62$ are coded by the size of the 
symbols: They are small for $-0.4 < [M/H]_1 < +0.2$ 
and large outside this range. The positive metallicities ($[M/H]_1 > 0$) 
are shown by filled symbols and negative metallicities
($[M/H]_1 \le 0$)  are shown by open symbols. 
Square symbols are used for binaries with velocities taken from 
\citet{bilir2005}. Velocities for binaries without reliable
$[M/H]_1$ are marked by crosses. Velocities for the red,
short-period binary CC~Com (12:2) are  marked by 
the right-directed arrows. For the high-velocity 
binary FU~Dra (3:5), the velocity component $V = -119.4$ km~s$^{-1}$ 
is outside the coordinate box.
}
\label{fig_kinem_Fe}
\end{center}
\end{figure}

After the pioneering investigation by \citet{gb88}, the only
study of kinematic properties of the EW binary stars was
by \citet{bilir2005} who 
considered all short-period binaries including the EW systems.  
It utilized results on the center-of-mass velocities,
$V_0$, obtained during the DDO program as published 
in the papers DDO-1 to DDO-9, so that results from the six subsequent
DDO papers (Table~\ref{tab_papers}) were not included; instead the study
used older, inhomogeneous data from various sources. 
In this paper we purposefully avoid diversity 
of sources and determine space velocities only for the EW binaries
which have proper motions based on the new Hipparcos reductions  
\citep{leeu2007}\footnote{The important, extreme short-period, 
red binary CC~Comae (which had not been observed by
Hipparcos) was included  using the data of \citep{Klem1977}. 
Its kinematic properties are similar to those of the F-subsample stars.} 
and the center-of-mass radial velocities 
$V_0$ determined during the entire DDO survey. 
The input data are listed in Table~\ref{tab_kinem1}.

\placetable{tab_kinem1}         

\placetable{tab_kinem2}         

The calculated $U$, $V$, $W$ space
velocity components, as listed in Table~\ref{tab_kinem2}, 
are corrected for the solar motion
\citep{Cos2011} relative to the Local Standard of Rest (LSR),
with $U$ counted positive toward the Galactic center,
i.e.\ using the same convention as \citet{bilir2005}.
We found that median differences between that paper and our 
determinations reflect only the difference in the treatment of the LSR
correction. We brought the Bilir et al.\ velocities to 
the LSR-corrected system of velocities by adding the solar velocity 
correction, $[+8.5, +13.4, +6.5]$ km~s$^{-1}$ as determined by
\citet{Cos2011}. After the adjustment,
the median values for the common stars
differed by only $[-1.6, +0.4, -0.3]$ km~s$^{-1}$
which mostly reflected the different assumed distances to the binaries.
The median error for our determinations of velocity
components is 1.9 km~s$^{-1}$, which is identical
to that in Bilir et al. However,
individual velocities sometimes differ quite a lot and
our error estimates generally tend to be larger. The largest differences
appeared for FG~Hya (1:6), $U = +70.1$ km~s$^{-1}$ for Bilir et al.\
and $U = 41.9$ km~s$^{-1}$ in our case, and for V2357~Oph (8:5),
$[V, W] = [-56.8, -37.0]$ km~s$^{-1}$ for  Bilir et al.\
and  $[V, W] = [-93.7, -57.7]$ km~s$^{-1}$ for our determinations.
These large discrepancies result from two sources: (1)~our use
of exclusively \citet{leeu2007} proper motions with re-determined errors, 
and (2)~larger uncertainties in our assumed
distances, which are hopefully more realistic. Thus,
we used mostly our velocities, but -- when not
available because of the more restrictive criteria in the reductions of
\citet{leeu2007} -- we added Bilir et al.\
determinations. The latter are listed in Table~\ref{tab_kinem2} 
in square brackets.
The number of new velocity determinations is 65 with 
an additional 13 velocities from the Bilir et al.; among those
78, 50 binaries belong to the F-type sub-sample. 

Figure~\ref{fig_kinem_Fe} shows that space velocities 
of EW binaries are rather typical for
the solar neighborhood. We discuss the results in terms 
of the velocity dispersions below. Here we note that 
only one binary has a large
velocity: For FU~Dra (3:5) the space velocity vector is
$[-99.3,  -127.2,  +113.4]$ km~s$^{-1}$, but the
errors are large, about $\pm 25$ km~s$^{-1}$ for each
component. The full space
velocity of the star is 197 km~s$^{-1}$, which sets the binary 
apart from the rest of our sample. 
Our determination gives a low metallicity of FU~Dra,  
$[M/H]_1 = -0.58 \pm 0.19$, but there exists no Str\"{o}mgren
photometry to support this value. This binary may be the only 
{\it thick disk\/} representative in our sample.

Besides FU~Dra, there are six other F-type stars with velocities 
$V_{\rm space} > 60$ km~s$^{-1}$ (see Figure~\ref{fig_vels_Fe}): 
CK~Boo (2:2): [82, $-$ , $-0.33$], 
GR~Vir (2:9): [68, $-0.05$, $0.00$],
OU~Ser (3:10): [103, $-0.39$, $+0.05$], 
GM~Dra (6:5): [61, $-$, $-0.50$],
and HN~UMa (8:8): [80, $-$, +0.41], 
where the first number after the star's 
identification is the spatial velocity in 
km~s$^{-1}$ while the second and third numbers are 
$[M/H]_{\rm m}$ and $[M/H]_1$.
These binaries may show some tendency for negative metallicities,
but the data are not accurate enough for such a conclusion.

The scatter of velocities as shown in Figure~\ref{fig_kinem_Fe} 
gives us some indication of the typical age of the
EW sample. The velocities increase  for later spectral 
types within the F-type sub-sample which is typical for 
the solar neighborhood.
The formal dispersions of velocity components for 43
F-type binaries (excluding FU~Dra) are: 
$\sigma U = 32.2$ km~s$^{-1}$, $\sigma V = 22.6$ km~s$^{-1}$
and $\sigma W = 13.1$ km~s$^{-1}$. 
When the sample is enlarged to 49 objects by adding slightly 
less secure velocities from \citet{bilir2005}, the dispersions 
are:
$\sigma U = 32.7$ km~s$^{-1}$, $\sigma V = 22.7$ km~s$^{-1}$
and $\sigma W = 14.4$ km~s$^{-1}$. When compared with 
the results of \citet{holm2009}, the 
velocity dispersions suggest the {\it thin disk\/} population 
with the age of about 3.0 -- 5.5 Gyr.

\begin{figure}[ht]
\begin{center}
\includegraphics[width=11.0cm]{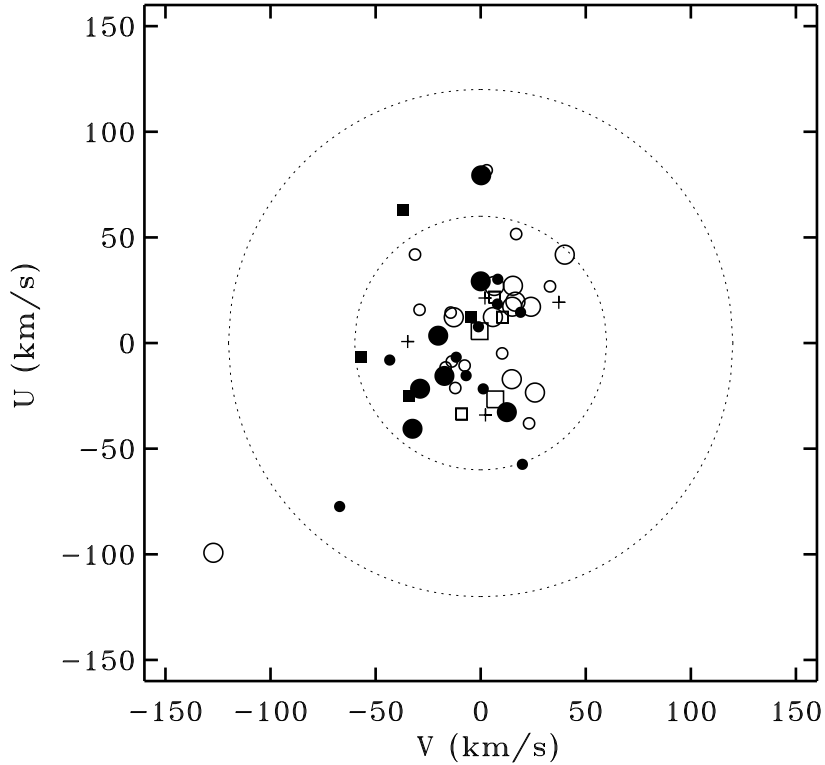}    
\caption{
\small 
The galactic disk-plane components of the space velocity, 
$U$ and $V$, for the F-type sub-sample are shown here 
with metallicities coded by the size of the 
symbols: small for $-0.4 < [M/H]_1 < +0.1$ and large outside
this range, with positive metallicities 
shown by filled symbols and negative metallicities 
shown by open symbols. Squares are for binaries 
with velocities taken from \citet{bilir2005}.
Crosses mark binaries with available velocities but 
without determined metallicities. 
The dotted-line circles have radii of 60 and 120
km~s$^{-1}$. This figure can be directly compared with 
Figure \ref{fig_B-V_dbv} in \citet{gb88}. 
}
\label{fig_kinem_uv}
\end{center}
\end{figure}

\begin{figure}[ht]
\begin{center}
\includegraphics[width=9.0cm]{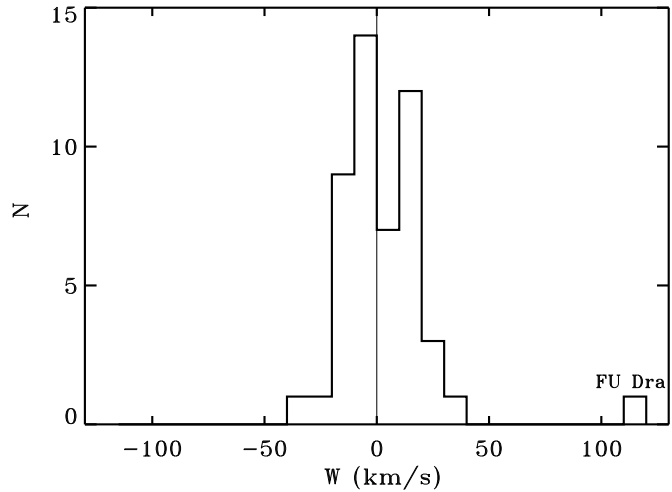}    
\caption{
\small 
The histogram of $W$ velocity components (perpendicular
to the galactic disk) for the F-type binaries.
OO~Aql has been excluded from the sample.
}
\label{fig_histW}
\end{center}
\end{figure}

The galactic-plane velocity components $U$ and $V$ for 
the F-type subsample 
are shown in Figure~\ref{fig_kinem_uv}.
The low and high metallicity objects do not separate
strongly in the two velocity components; only FU~Dra
is set apart from the rest of the EW binaries. 
The absence of an obvious correlation between 
space velocities and metallicities 
can be also seen in Figure~\ref{fig_vels_Fe}. 

The special position of FU~Dra is particularly strongly
visible in the $W$-component velocities (the lower left
panel of Figure~\ref{fig_kinem_Fe}) and in their 
histogram (Figure~\ref{fig_histW}). 
The $W$ velocity dispersion is an useful indicator of 
the population: For the solar neighborhood,
the characteristic values are $\sigma W \simeq 20$ km~s$^{-1}$ 
for the thin disk and 45 km~s$^{-1}$ for the thick disk
\citep{Edvard1993,Veltz2008}. Thus, the small scatter
in the $W$ velocities in Figure~\ref{fig_histW} is 
indicative of small velocity errors, reliable distances, and
strongly suggests that the analyzed
EW binaries belong to the thin-disk population.

\section{Metallicities and evolution}
\label{evolution}

The prevailing view on how EW binaries form and evolve
sees them as a result of the angular-momentum-loss
driven evolution of close but detached binaries.
The original periods of these binaries are expected to be about
2.5 -- 3.5 days; they are rather mildly evolved objects when they 
enter into mass-exchange and/or mass-loss phase.  
It is not clear if during the contact-binary 
formation the mass-ratio reverses, 
but many indications and theoretical 
predictions suggest this to be the case \citep{step2006,step2009}. 
An event of mass-reversal would expose the 
interior of the mass-losing companion, but we 
would not see any changes in our spectral window. Besides, 
most of the binary components are not sufficiently
massive for the CNO cycle and an interruption of the 
p-p cycle and exposure of the interior material
will not alter the observable metallicities.
Therefore, the metallicities that we measured reflect properties
of EW progenitors and are not expected to be
modified by their subsequent evolution.

\begin{figure}[ht]
\begin{center}
\includegraphics[width=11.5cm]{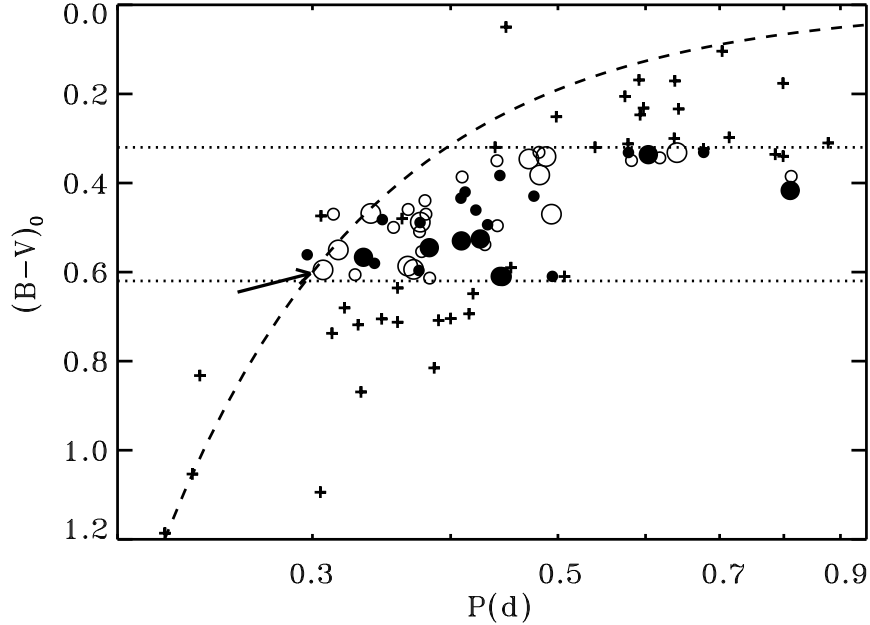}    
\caption{
\small 
The period -- color relation for the EW binaries analyzed
in this paper with metallicities $[M/H]_1$ coded by circles for the 
F-type sub-sample: The sizes of the circles are small for 
$-0.4 < [M/H]_1 < +0.2$ and large outside
this range, with positive metallicities shown by filled symbols
and negative metallicities shown by open symbols.
The stars not in the F-subsample or with unreliable
determinations of metallicities are marked by crosses.
The two detached
binaries have been removed from this plot which
applies to contact binaries only. The figure uses the same
convention as in the papers of \citet{eggen1961,eggen1967}
with the color axis directed downwards and with the logarithmic 
scale for the period.  FU~Dra is pointed out with an arrow. 
The broken line 
gives the ``Short-Period Blue Envelope'', as discussed in the text. 
}
\label{fig_Per_Fe}
\end{center}
\end{figure}

The evolutionary state of a EW binary requires some assumptions 
on the energy-exchange and mass-exchange model. While the 
Lucy model \citep{Lucy1968a,Lucy1968b} is the most popular
model for EW binaries, it must still be the subject of continuing 
scrutiny: It provides excellent reproduction of light curves
(which however have low information content) but the spectroscopic
support is still rather fragmentary. The strict agreement 
with the Lucy model has been recently questioned for the particularly
important and well-studied low mass-ratio
binary AW~UMa \citep{PR2008}.
On the other hand, very strong interaction 
between the components cannot be questioned because
we do see an excellent equalization of temperatures over
the whole binary structure. Irrespective of the detailed model, 
much can be said about the state of these binaries by simply
using the essential observational quantities: 
(1)~the period, $P$, which is known practically without any error, 
(2)~the mass-ratio, $q = M_2/M_1 \le 1$, 
known from radial-velocity determinations with uncertainties 
typically at the level of $\epsilon q \le 0.01$, 
and (3)~the $(B-V)_0$ color which is the least reliably 
known, but a crucial quantity, as it can be used to
relate the above two quantities to the thermal state of the stars.
To this set, we can now add two quantities which correlate
with the age, the metallicity $[M/H]_1$ and the space velocity
vector $[U,V,W]$\footnote{We avoid using   
individual masses  obtained from radial velocity 
semi-amplitudes $K_i$ because they are geometric-model
dependent through the orbital inclination.}.

The period -- color relation in Figure~\ref{fig_Per_Fe}
summarizes the essential properties of the EW binaries 
discussed in this paper in terms of periods, colors and metallicities. 
The relation was discovered by \citet{eggen1961,eggen1967}
and has been used since as one of the tools to infer the
evolutionary state of EW binaries. Its existence has been
the strongest argument for inferring that the EW 
binaries generally follow the Main Sequence with 
the color being a useful proxy for the mass of the primary 
component. 
The $[M/H]_1$ metallicity determinations of the F-type sub-sample 
(between the two dotted horizontal lines) are coded 
in Figure~\ref{fig_Per_Fe}
by open and filled circles of different size.
The broken line in the figure gives the ``Short-Period Blue Envelope'' 
(SPBE) as defined in \citet{ruc1998b}. 
Stars close to the SPBE are expected
to be either (1)~un-evolved, without any evolutionary expansion, 
or (2)~of low-metallicity, because of the smaller sizes of such 
binaries and the reduced metal blanketing in the $B$-band
\citep{ruc2000}. 
We know that the EW binaries cannot be young and un-evolved
because they do not appear in young clusters, so the 
low-metallicity assumption for systems close to the SPBE
is more likely.
We see in Figure~\ref{fig_Per_Fe} that the metal-poor  
binaries show some tendency to be closer to the SPBE than those
with high metallicities. The high-velocity,
low-metallicity binary FU~Dra is exactly on 
the SPBE line. In general, however, the separation in metallicities
is not very strongly visible in this figure. 

\begin{figure}[ht]
\begin{center}
\includegraphics[width=11.5cm]{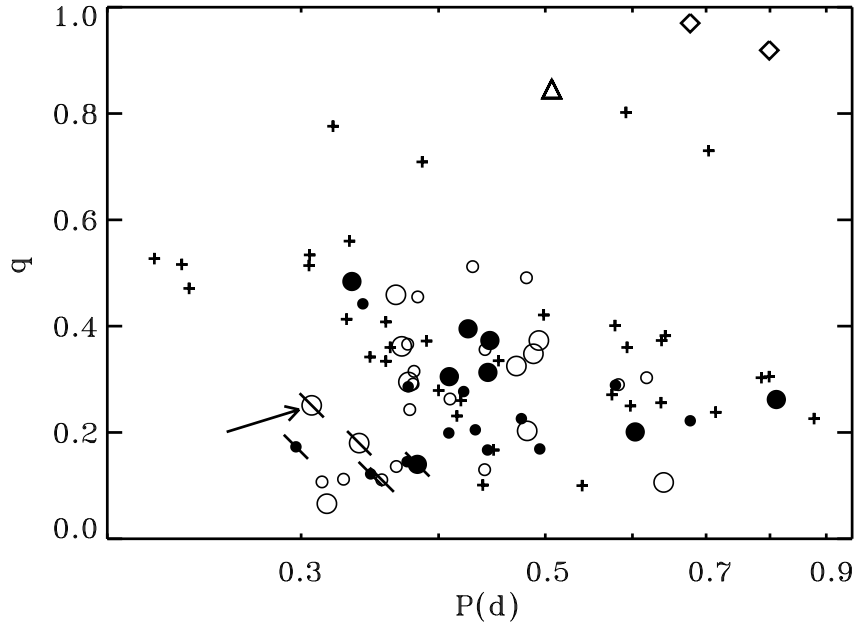}    
\caption{
\small 
The spectroscopic mass-ratios $q$ are shown versus the orbital
 period $P$ (in days). 
The F sub-sample binaries with reliable determinations
of $[M/H]_1$ are marked by circles with small sizes for 
$-0.4 < [M/H]_1 < +0.2$ and large outside
this range, with positive metallicities shown by filled symbols
and negative metallicities shown by open symbols.
Binaries with spatial velocities $>60$ km~s$^{-1}$
are marked by slanted bars and FU~Dra is additionally pointed out with
an arrow. 
The stars outside the F-sub-sample range and with unreliable
determinations of metallicities are marked by crosses.
The two detached binaries in the upper right corner are marked by
rhombuses while the unique contact binary OO~Aql is marked by a triangle.  
}
\label{fig_B-V_Fe3_qq1}
\end{center}
\end{figure}

Figure~\ref{fig_B-V_Fe3_qq1} shows the
relation between the orbital period and the spectroscopic  mass ratio.
This relation has already been shown before \citep{ruc2010a,ruc2010b} 
to demonstrate an unexpected bifurcation in the mass-ratio 
of contact binaries observed in the DDO survey:
For periods longer than $P \simeq 0.35$ days, the mass ratio 
can be either small, $q < 0.5$, and decreasing with the period, 
or large, $q > 0.7$. There is a prominent gap between the two branches. 
The DDO survey was not biased in any way against the
mass ratio so that this result is well established.

The upper branch in Figure~\ref{fig_B-V_Fe3_qq1} is defined by 
atypical binaries: The two detached binaries (V753~Mon and V1130~Tau) 
appear there as expected, but among genuine EW binaries we
see only those with the earliest or the latest spectral types. 
The early-type ones are V2150~Cyg (4:3) and V449~Aur (10:3)
with the spectral types estimated as A5/6V and A5/6V (or A2IV).
The late-type are EW binaries with peculiarities: Two
binaries with masses typical for the F spectral type
($>2 \,M_\odot$) yet with late spectral types,
SW~Lac (10:5) and OO~Aql (12:1) (the latter additionally
with highly uncertain reddening, see Section~\ref{reddening})
and AU~Ser (14:9) showing a very large, single, dark 
spot\footnote{Large spots are easily detectable in BF profiles as well defined,
migrating ``notches'', but they have not been 
detected in spite of extensive scrutiny
of thousands of BF profiles for all EW binaries of the
DDO program; AU~Ser is an exception. Large spots which are
invoked frequently to explain systematic discrepancies in 
photometric analyses of light curves may be an indication of 
a deficiency of the contact model \citep{prib2011}. 
However, small, localized spots were seen in high-quality 
data for AE~Phe \citep{Barnes2004} and AW~UMa \citep{PR2008}.}. 

The better populated, 
lower branch in Figure~\ref{fig_B-V_Fe3_qq1} appears to 
contain genuine contact binaries. \citet{GN2006}
pointed out that the less massive components of binaries 
located along this branch have very similar masses of 
about $0.5\,M_\odot$. 
The lower branch contains several F-type binaries with reliable 
determinations of metallicities $[M/H]_1$. 
Of special note are the shortest-period F-type binaries  
with orbital periods around $P \simeq 0.3 - 0.35$ days,
particularly a group with
large spatial velocities and mostly low metallicities, 
hence probably older than the rest. They all 
have small mass ratios, $q \simeq 0.1 - 0.2$.
In contrast, the remaining F-type sub-sample binaries, presumably
younger and/or less evolved, have larger mass ratios,
$q \simeq 0.3 - 0.6$, and a wide range in metallicities.
This mass-ratio spread at the shortest periods
may be the best defined manifestation of evolution
for EW binaries. 
Evolution towards small mass ratios was predicted from
considerations of the stability of Lucy's 
model \citep{Lucy1968a,Lucy1968b} in \citet{Lucy1976}.
Although we see the predicted progression in the mass ratio 
(the vertical direction in Figure~\ref{fig_B-V_Fe3_qq1}),  
we cannot exclude a systematic evolution to progressively 
shorter periods at a constant mass-ratio.

No clear indication of evolution is provided by variations
of orbital periods of EW binaries. The period variability is known 
to be strong and easily detectable \citep{Kreiner1977}. 
In many cases it is relatively regular and induced by 
a companion as they are very
common for EW binaries \citep{triples1,triples2,triples3}.
However, it appears that the period variations must primarily reflect 
mass-transfer effects within the binary itself and not evolutionary
effects. As shown in the study of period variations
based on the OGLE survey \citep{Kubiak2006}, 
the period time derivative is equally likely
be positive as negative and its size is typically large in the 
absolute sense, implying variations in time-scales of about 
$1.5 \times 10^6$ years. Such a rapid variability
is simply too fast to result from evolutionary changes;
it is at least some 10 times faster 
than the thermal time scale of the more massive component
which is expected to determine the period evolution rate.  
These reasons, and the strong inhomogeneity of the 
period-variability data for our sample of 90 binaries, 
led us to regard the period changes 
as not related to their evolutionary state of the 
analyzed EW binaries.

\section{Discussion and conclusions}
\label{conclusions}

\subsection{Summary of the results}

This paper presents a first attempt to determine the metallicities 
of W~UMa-type systems from the spectroscopic data.
A novel technique based on the broadening function method
calibrated using the synthetic spectra was developed; it 
determines the integrated strengths of metallic 
lines within a set spectral window. 
Several implementation problems have been encountered; 
possible methods of circumventing them in the future 
are discussed in the next Section~\ref{recommend}.

The most reliable metallicity determinations were 
obtained for the ``F-type sub-sample'' of 
early-F to early-G type systems with $0.32 < (B-V)_0 < 0.62$
consisting of more than one half of all binaries of the total sample 
of 90 objects. 
Our results are shown in the graphical form in 
Figure~\ref{fig_B-V_logS} (the raw determinations $[M/H]$)
and in Figure~\ref{fig_B-V_Fe3} (the determinations corrected 
for a trend inherent to the method, $[M/H]_1$):
The F-type EW binaries have roughly solar metallicities, but with a
relatively large scatter which is partly due to the large
determination errors. 
The spectroscopically determined metallicities, $[M/H]_1$, correlate 
with the Str\"{o}mgren photometry $m_1$-based metallicities $[M/H]_{\rm m}$
supporting the validity of both methods. The spectroscopic results 
strongly suggest that the $m_1$ index is a valid measure of metallicity 
and (for the F-type EW binaries) is not affected by 
chromospheric activity as was suspected before. 

The distribution of $[M/H]_1$ for the F-type sub-sample
extends between $-0.65 < [M/H]_1 < +0.5$ 
(Figure~\ref{fig_vels_Fe}),
with half of all systems having metallicities
within $-0.37 < [M/H]_1 < +0.10$. 
The distribution is asymmetric, with the mean at $-0.1$,
the median at $-0.04$ and a tail extending towards low metallicities. 
The determination errors are estimated at about 
$\epsilon [M/H] \simeq 0.16$ and they contribute to the width
of the distribution. While the preferentially negative values
of the metallicities is entirely consistent with a typical 
stellar population in the solar neighborhood,
the distribution may be biased
against positive metallicities.
This is due to the line saturation effects and our removal of 
metallicities which required extrapolation 
beyond the largest model value of $[M/H] = +0.5$. 
It must be stressed that the absolute level (zero point) of our
determinations is anchored to the $m_1$ calibration 
and is subject to any systematic error of the former.

We determined and analyzed the kinematic data for 50 binaries
of the F-type subsample.   
We found that all but one (FU~Dra) show small to moderate 
velocity dispersions in all three principal directions.
The dispersion is particularly small in the
direction perpendicular to the galactic disk,
$\sigma W \simeq 14$ km~s$^{-1}$ which is a
characteristic property of the solar-neighborhood, thin-disk 
population with a typical age of about 3.0 -- 5.5 Gyr. 
The high-velocity binary FU~Dra with
$V_{\rm tot} = 197$ km~s$^{-1}$, $[M/H]_1 = -0.58 \pm 0.19$,
shows $W = 113$ km~s$^{-1}$. This may be 
the only thick-disk star in the sample or an object which
experienced a strong out-of-plane scattering event in the past.

The findings concerning the F-type EW binaries lead to refinement
of the conclusions of \citet{gb88}: While
it is confirmed that the EW binaries are rather old objects,
they do not seem to differ from most typical, thin disk dwarfs 
in terms of their kinematic properties. 
The previous result used less accurate proper-motion data 
and included late-type EW binaries 
which -- similarly to other solar neighborhood
stars -- tend to have larger velocities than F-type stars.
With the exception of FU~Dra which is moderately metal
poor, we see no obvious correlation between metallicity
and spatial velocity for 44 F-type sub-sample binaries with the new
velocities and reliable determinations of $[M/H]_1$.

While the mean metallicities give almost no constraints 
on the ages, as the age -- metallicity relation allows a range 
from the present to 10 Gyr \citep{Reid2007},  
the ages estimated from the kinematics 
(3.0 -- 5.5 Gyr) are in agreement with the upper limit 
for thin-disk stars not exceeding 7 Gyr 
as estimated from kinematic models 
\citep{Schon2009}\footnote{\citet{FB2009} quote 
observational estimates on {\it a lower limit\/} 
to the thin disk age of 9 Gyr. This issue appears to be
far from being settled.}. Because the
EW binaries appear in large
numbers in old open clusters, for example in NGC~6791 
\citep{Moch2002,Moch2005} with an age
estimated at 7.7 -- 9.0 Gyr \citep{Grund2008},
and in still much older globular clusters \citep{ruc2000}, they
must keep on forming during the lifetimes of these clusters. 
This explains the rising frequency of
EW binaries with age for old open clusters \citep{KR93,ruc1998b}
as having its source in the steeply increasing numbers of stars 
along the MS luminosity function rather than in the 
long survival times of the EW binaries. 
 
The color (Figure~\ref{fig_Per_Fe}) and the mass-ratio 
(Figure~\ref{fig_B-V_Fe3_qq1}) dependences on the orbital period
give interesting evidence on the evolutionary state of the 
analyzed EW binaries. 
Here, we see a group of binaries with relatively short periods 
($0.3< P < 0.4$ days) and low mass ratios ($0.05< q < 0.3$) 
which have a tendency for large spatial velocities 
($>60$ km~s$^{-1}$) and low-metallicities.
Thus, they appear to be old stars. 
Although we cannot say if these binaries 
evolved by drifting to small mass-ratios (as predicted by
Lucy's contact model) or to short orbital
periods through loss of angular momentum, 
the special location of these binaries in 
Figure~\ref{fig_B-V_Fe3_qq1} is striking 
and requires further research.  

Finally, we note that three triple systems with faint
companions, $L_3/L_{12} < 0.05$,
did not show drastic deviations from the other results,
so very common triple systems with faint companions
may be (cautiously) included in similar investigations.
Also, the two equal-mass, detached binaries 
appeared to merge in between the genuine early-type contact binaries, but
we could not contribute any firm results on their metallicities 
for these binaries because they fell outside of the best studied
sub-sample of F-type binaries. The location of the detached binaries
among the upper mass-ratio branch in Figure~\ref{fig_B-V_Fe3_qq1} may help
in identifying pre-contact binaries or binaries  masquerading to be
in contact.

\subsection{Recommendations for the future}
\label{recommend}

The spectroscopic metallicities were determined 
using the combined strengths of
metallic lines within the spectral window 5080 \AA\ to 5285 \AA\
which contains the Mg~I triplet. 
For early F-type EW binaries, for which our metallicity 
determinations are the most secure, the lines of the 
triplet produce about 10\% of the total line strength within 
the window, so that mostly iron lines
contributed to the measured line strength. However, 
the Mg~I triplet contribution increases rapidly 
for later spectral types; with it, increases the dangers of
contamination by chromospheric emission. 
Thus, the Mg~I triplet window, which served well for radial velocity
determinations, is not the optimum region for line strength
measurements. Less congested spectral windows, such as
the region of the Mg~II 4481 \AA\ line 
for spectral types B9V to A5V or
the region around 6290 \AA\ for late-type stars may give better
results.

The pseudo-continuum placement was 
identified as the main limiting factor in our method.
In the future, it will be essential to pay particular attention to the
rectification step, perhaps by using nightly observations
of smooth-continuum spectrophotometric standard stars or
synthetic spectra calculated with and without the lines. 
When better temperature information for
EW binaries becomes available (e.g.\ from infrared colors), 
an additional improvement could result from using model 
template spectra computed for individual binaries. 

A systematic photometric survey of the W~UMa-type binaries
in the Str\"{o}mgren $uvby$ system is very much needed. 
Our spectroscopic results are in very good agreement with
metallicities derived using the $m_1$ index data which 
require relatively lower observational expenditures than 
a spectral investigation. 
Finally, it would be interesting to repeat this work when new 
astrometric and spectroscopic data from the GAIA mission 
\citep{lind2008} and the Gaia-ESO Public Spectroscopic Survey
\citep{gil2012} become available.

\begin{acknowledgements}

The authors would like to express their thanks 
to the DDO Telescope Assistants: Jim Thomson,
Heide DeBond and Wen Lu for many many hours of their work at the DDO telescope.
Jim worked practically all his life in the David Dunlap Observatory. 

Thanks are due to Dr.\ Kazimierz Stepien for his comments on the
early version of the paper and to Drs.\ Edward Guinan, Petr Harmanec and
Steven Shore (the participants of the Hvar, Croatia meeting on binary stars 
in July 2012) for pointed questions. 

The Natural Sciences and Engineering Research Council of
Canada has supported the research of S.M.R.\ over the duration
of the DDO close-binary program; several visits of collaborators who
contributed to the DDO series were supported from these funds. 
T.P.\ acknowledges support from the EU in the FP6~MC~ToK 
project MTKD-CT-2006-042514.
This work has also partially been supported by VEGA project 2/0094/11
and by the Slovak Research and Development Agency under the contract
No. APVV-0158-11. The research made use of the SIMBAD database, 
operated at the CDS, Strasbourg, France.

\end{acknowledgements}

\appendix
\section{Uncertainties and the error budget}
\label{appendix-A}

\subsection{Pseudo-continuum and rectification}
\label{continuum}

Most EW binaries in the sample are cooler 
than the template HD~128167 ($\sigma$ Boo, F2V, $B-V=0.36$) 
so that the values of $S$ are slightly larger 
than unity with the observed range $-0.2 < \log S < +0.5$. 
The measurements of $S$ show a very small orbital-phase scatter.
The median random measurement error computed from the
orbital phase scatter of $\log S$ is $\epsilon_1  = 0.0032$, with
the smallest error in the sample of 0.0006 and the largest of 0.020. 
We describe below that these very small formal errors are
not representative of the actual, much larger uncertainties.

\begin{figure}[ht]
\begin{center}
\includegraphics[width=11.5cm]{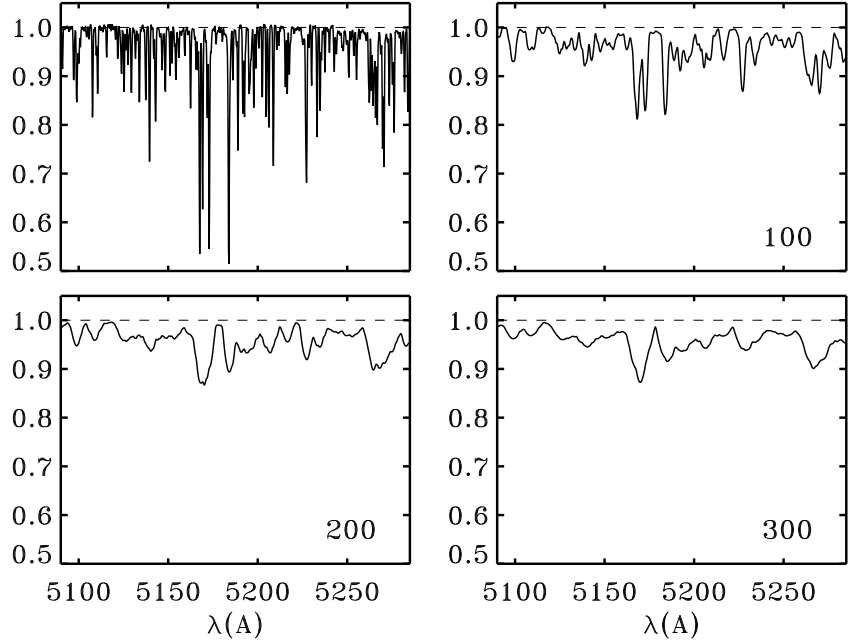}    
\caption{
\small 
Broadening of the spectrum due to rotation. 
The upper left panel shows the template spectrum while the
remaining panes give the template spectrum convolved with
the theoretical rotational profiles of single stars 
for three values of $V \sin i$,
100, 200 and 300 km~s$^{-1}$, as given in each of the panels. 
For the EW binaries the line broadening is complicated by
additional phase-dependent blending.
}

\label{fig_Vsini}
\end{center}
\end{figure}

The heavy velocity broadening and blending of late-type spectra
caused by rapid rotation profoundly change their appearance, 
as can be seen in Figure~\ref{fig_Vsini}. The figure
illustrates a simpler case of rapidly rotating 
{\it single stars\/} at the $V \sin i$ values typically observed
for components of EW binaries. 
One should note the strongly decreased depths and large 
depressions in the pseudo-continuum
between the spectral lines. For EW binaries, the additional
complication is the rapidly variable line blending 
in time scales as short as a few minutes. 
These rapid spectral variations have strong effects on observed
pseudo-continuum levels with departures 
from the correct continuum level leading to changes in $S$. A
downward displacement by $\Delta$ (see Fig.~\ref{fig_BF})
from the real continuum level to a pseudo-continuum level
will cut off weak lines and will reduce $S$
by the factor $(1 - \Delta)$. 

For an idealized case of a uniform (across the spectral window)
vertical shift and the spectrally
matching template, the Broadening
Function technique has the potential to permit evaluation of 
$\Delta$ from the BF baseline displacement, $\delta$: When
$\delta = 0$, the spectral continuum is correctly 
positioned; for any $\delta \ne 0$, the integral of the BF 
above the baseline requires linear scaling by 
$1+\Delta$ with $\Delta = C \times \delta$, with $C$ being
a factor converting linear units of the spectrum height into 
the units of the $S$ integral. Unfortunately, 
such an approach could not be used here because of 
uncertainties at the rectification step of the spectral processing due
to absence of information on location of the actual continuum level.
Rectification hides -- in its simplicity -- many complex factors.
For lack of real information, it is usually done by joining the highest
points of the pseudo-continuum. 
This not only corrects for the genuine depression of the stellar
continuum due to the line blending, but 
mainly removes many instrumental problems affecting the spectra,
such as chromatic efficiency variations in the spectrograph
(the grating efficiency combined with the detector sensitivity
non-uniformities), positioning of the object on the
spectrograph slit, chromatic differential refraction, etc.
These contributions are not uniform across the spectral window.
Thus, $\delta = 0$ is not a guarantee that the proper level 
of the spectral continuum has been established.

\begin{figure}[ht]
\begin{center}
\includegraphics[width=11.5cm]{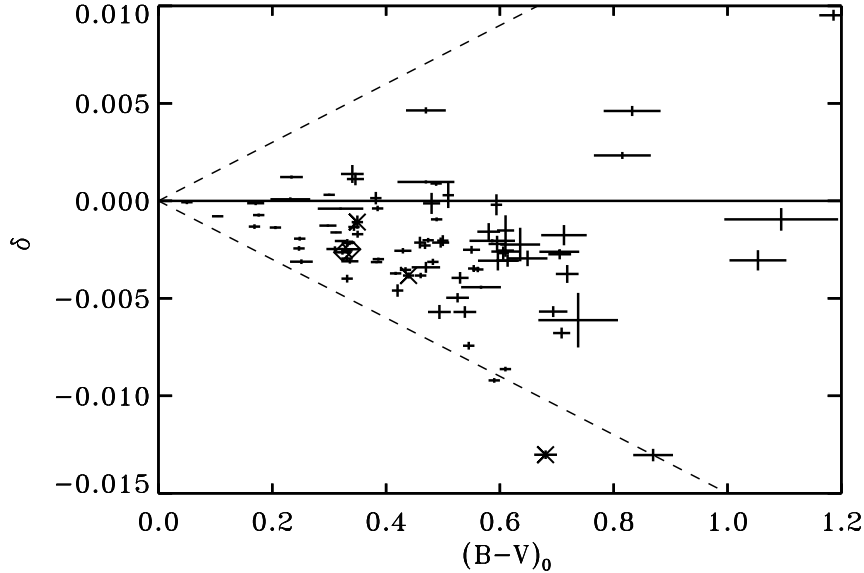}    
\caption{
\small 
The measured values of the BF baseline shifts $\delta$ with their
formal measurement errors shown by vertical error bars. The 
$(B-V)_0$ colors (Section~\ref{efftemp})
with their uncertainties are shown by horizontal bars. 
The broken lines give the adopted maximum range of $\delta$ 
for estimates of the $\log S$ dominant systematic errors
$\epsilon_2$ (Section~\ref{errors}).  
Two detached binaries are marked by the over-plotted diamonds
while three triple systems with $L_3/L_{12}<0.05$ are marked
by the slanted cross (X) symbols. 
}
\label{fig_B-V_delta}
\end{center}
\end{figure}

Uncertainties in the rectification did not substantially affect 
the results of the DDO radial velocity program because its 
influence on {\it positions\/} of Broadening Function
peaks was small. However, for the {\it line strength\/} 
determination, the roles of rectification and
the pseudo-continuum positioning are essential. The 
spectra do not provide any information the the pseudo-continuum 
shifts nor on errors in
approximating pseudo-continua by broken lines; these shifts
do not show up in the scatter of the measured values
of the BF integrals $S$ nor in the scatter of 
the $\delta$ shifts. In fact, the $\delta$ shifts
for individual objects were found to be surprisingly similar 
with very small formal uncertainties. 
This was unexpected, but is explainable 
by the way the DDO spectra were reduced:
Each star was analyzed by one investigator within typically hours, 
at most days; 
he or she tended to use the same choices in the piece-wise
approximation when joining the highest points of the pseudo-continuum. 
This resulted in similar {\it systematic\/}
errors for all spectra of a given star. Thus, the values of
average $\delta$ and of their scatter are almost meaningless for
individual objects. 
However, the values of $\delta$ for the whole sample,
by providing information of their spread between stars, 
can be still useful in a statistical sense:
The observed {\it range\/} of $\delta$ can be used to estimate 
the expected {\it range\/} of $\Delta$ and thus the uncertainties 
of the integrals $S$ entering through the corrections $(1 + \Delta)$.

The distribution of the individual values of $\delta$ versus the
star colors is shown in Figure~\ref{fig_B-V_delta}; we use there the
de-reddened colors as these represent spectral complexity. 
The errors in the shifts $\delta$ are not the same across the 
whole range of effective temperatures: 
They are small for early-type binaries having spectra with few, 
weak lines, and are large for late-type spectra which even 
without rotational and orbital broadening show strong 
blending. There is
a virtual absence of $\delta$ shifts for $B-V < 0.25$ and
a rapid increase in their size for red stars. 
Most $\delta$ shifts are negative implying too low 
measured lines strengths for later spectral types. 
Note however the presence of a few {\it positive\/} values of
$\delta$ indicating that 
the continuum was placed too high and that the shift was over-corrected.

The lines in Figure~\ref{fig_B-V_delta} give our adopted, rather
generous estimate of the maximum range of $\delta$ 
as a function of the de-reddened color $(B-V)_0$: 
$\delta_{\rm max} = \pm 0.015 \, \max(0, (B-V)_0-0.1)$.
The measured values of $\log S$ were corrected using: 
$\log S_{\rm corr} = \log S_{\rm meas} + \log(1 + C \delta)$.
Although only in the idealized case of
perfect rectification the two shifts would be
proportional, $\Delta \simeq C \delta$, we
nevertheless assumed such scaling. The 
 multiplier $C$ depends on the shape of the BF.
Experiments with various shapes of Broadening Functions
have led to $C = 12 \pm 3$. 
From now on, we will call $\log S_{\rm corr}$ simply $\log S$
remembering that these values include the baseline level
corrections. The corrections are not large; the largest 
for the sample was $|\log(1 + C \delta)|=0.06$.

The systematic error of $\log S$
due to the assumed scatter in $\delta$ is 
the second uncertainty in the evaluated values of BF integrals: 
$\epsilon_2 = (C+3) \, 0.015/2.3 \, \times \max(0, (B-V)_0-0.1)$; 
the coefficient $C$ has been increased by 3 to account for
its uncertainty.
This error was added in quadrature to the random measurement errors
$\epsilon_1$. As detailed in Appendix~\ref{errors},
$\epsilon_2$ dominates in the final uncertainty budget.

\subsection{Surface gravity and the synthetic 
$\mathbf{B}-\mathbf{V}$ colors}
\label{gravity}

We assumed that the gravities of most EW systems are approximately 
solar (i.e.\ $\log g_\odot = 4.44$ cgs) because the binaries are located
along the Main Sequence and our previous luminosity class estimates
were mostly consistent with the luminosity class V (dwarfs). 
The observed scatter in orbital periods
(for a given effective temperature) reaching $\Delta \log P \simeq 0.2$
\citep{RD1997} 
suggests that there exists a corresponding scatter in gravities, 
but probably not exceeding a factor
of about two, which is much less than the factor of 10 between 
the two model values of $\log g$ that we used. In the end, 
we used the differences in the theoretical values of $\log S$
between the two gravities, scaled to the difference
$\Delta \log g = 0.3$. 
The median differences in $\log S$ (over all metallicities) 
range within $[-0.035, 0]$. We give them in the last
column of Table~\ref{tab_models}. The absolute values of these
differences are identified as the third uncertainty of $\log S$.
This uncertainty, $\epsilon_3$, 
is entirely related to the use of the models
and not to the observational values of $S$.

The differences in $\Delta (B-V)$, between the observed
\citet{bess1979} and our theoretical values represent 
an inherent (systematic) 
uncertainty in the models which constitute the fourth source
of systematic error in $\log S$:
$\epsilon_4 = 
(\partial \log S_{\rm th}/\partial (B-V)) \times \Delta (B-V)$ .

\begin{figure}[ht]
\begin{center}
\includegraphics[width=9cm]{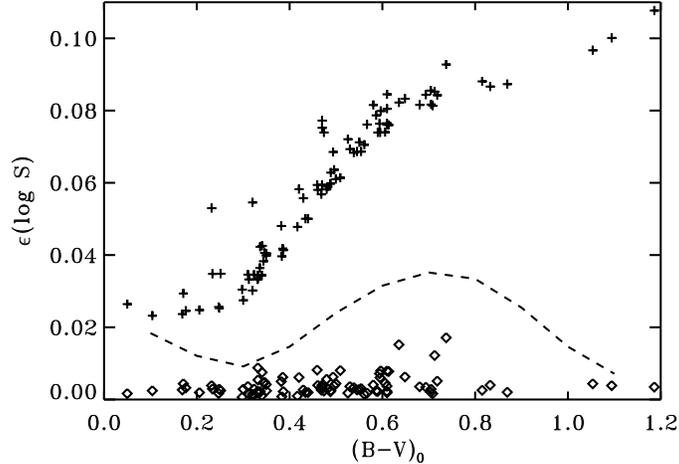}    
\caption{
\small 
The three main sources of $\log S$ errors versus $(B-V)_0$. 
The diamonds close to the lower margin 
show errors $\epsilon_1$ corresponding to the
orbital phase scatter in the measured values of $\log S$.
The crosses in the uppermost sequence show the combined errors
$\epsilon (\log S)$ consisting of all five components,
$\epsilon_1$ to $\epsilon_5$, all added in quadrature; they
are dominated by the systematic uncertainties 
in the continuum placement, $\epsilon_2$ (linearly
increasing with $B-V$, as described in the text)
and the unknown gravity, $\epsilon_3$ (broken line). 
}
\label{fig_logS_err}
\end{center}
\end{figure}

\subsection{Summary of uncertainty estimates}
\label{errors}

Estimates of the errors of $\log S$ and their dependence
on the color index are shown in Figure~\ref{fig_logS_err}.
In the summary below, the errors with index numbers 
all apply to $\log S$ and are non-dimensional, similarly to 
this quantity.
\begin{itemize}
  \item $\epsilon_1$ (Appendix~\ref{continuum}). It is the 
$rms$ error resulting from the orbital phase scatter in the
measured values of $\log S$. It is deceptively small:
the median is 0.0032, the maximum value 0.02. 
This is the only random error 
measured directly, the remaining ones listed below are
estimates of systematic errors. $\epsilon_1$
is shown along the lower axis of Figure~\ref{fig_logS_err}.
  \item $\epsilon_2$ (Appendix~\ref{continuum}). This
uncertainty represents difficulties
with continuum placement as they increase towards later
spectral types. This is the dominant source of
uncertainty. We assumed it to be linearly dependent on $B-V$; 
it reaches about 0.1 for $B-V=1.0$ 
(see Figure~\ref{fig_B-V_delta}). Although we assumed that
$\epsilon_2$ is symmetric around zero, the actual errors 
may have a skewed distribution with the dominance of negative
values.
  \item $\epsilon_3$ (Section~\ref{models}). This error 
is related to the use of the atmosphere models and
describes lack on information on the actual values of
gravity $\log g$. We used as error estimates 0.3 of the
median differences (for all metallicities) between the 
model values for $\log g=3.5$ and 4.5. 
This uncertainty reaches 0.035, as 
shown by the broken line in Figure~\ref{fig_logS_err}.
  \item $\epsilon_4$ (Section~\ref{models}). 
We found that our models 
differed systematically from the \citet{bess1979} calibration 
(Figure~\ref{fig_B-V_dbv}). The resulting uncertainty of $\log S$ 
is the result of projecting the calibration differences 
$\Delta (B-V)$ into the model dependence of $\log S$ on the color
index: $\epsilon_4 = 
(\partial \log S_{\rm th}/\partial (B-V)) \times \Delta (B-V)$.
For simplicity, we used the solar metallicity dependence 
for the derivative.
  \item $\epsilon_5$ (Section~\ref{models}). This is a similar
dependence to above, but it projects the
individual color uncertainties $\epsilon (B-V)$ (Section~\ref{bv})
into $\log S$. The same formula as above
for $\epsilon_4$ applies with the systematic trend $\Delta (B-V)$ replaced by
the color uncertainty $\epsilon (B-V)$.
  \item Three additional uncertainties in  
$\log S$ could not be characterized. They are expected to
produce a systematic trend in the measured line strengths $\log S$ 
making them systematically too low for late spectral types.
They results from: (1)~a skewed distribution of $\epsilon_2$ 
with dominant negative deviations (Appendix~\ref{continuum}), 
(2)~our use of a single template across the spectral
range and (3)~line emission filling-in the magnesium-triplet lines 
which -- if present -- would become more
important for late spectral types (Appendix~\ref{appendix-C}). 
We detected such a trend and removed it
using the Str\"{o}mgren photometry index $m_1$ 
(Sections~\ref{Strom} and \ref{results}).
\end{itemize}

\section{Is use of a single template justified?}
\label{appendix-B}

As described above, we used one template
spectrum of HD~128167 for determination of Broadening Functions
for all observations and for all atmosphere models. 
The template served as an intermediate device
for integration of information from the whole
spectral window into just one number characterizing the
total line strength for a given spectrum. This number
was then averaged for all spectra and 
interpolated in the pre-computed two-dimensional table of the
atmosphere-model BF strengths,  
$\log S = \log S(T_{\rm eff}, [M/H])$, to determine the metallicity 
$[M/H]$. This was a simple and convenient approach; it by-passed
the inherent problem of poorly known effective temperatures
of our targets and permitted following temperature uncertainties 
as they propagated into uncertainties of $[M/H]$. 

The weakness of the above approach is that
the BF method works perfectly well when the program and template
spectra are identical and were taken at the same spectral resolution
which controls how spectral lines are merged to define the 
pseudo-continuum. Because the spectra become richer at 
lower temperatures, more line blending occurs there even for sharp-line 
templates. This effect is missed when one template is used.
This will not show up in any BF baseline
shift ($\delta$) because it is the template which would require
modification, not the object spectra.
The correct approach would be to use an individual 
template spectrum for each object, computed for the
specific effective temperature and appropriately
adjusted in spectral resolution to that of the observed spectrum. 
We did not follow this approach
because we simply could not trust the available color measurements to 
provide correct effective-temperature estimates.
We did not have sufficiently enough
late-type template spectra to do tests
on the dependence of the results on the template used. Some
limited tests with another radial-velocity template that we
used frequently, HD~65583 (G8V), confirmed the expected tendency
for lowered values of $S$ when the template is too early. 

The assumption of a single template spectrum on the measured
values of $\log S$ has the same sign as the effect of 
the pseudo-continuum rectification and resulting 
negative $\delta$ shifts, 
both should produce too low $\log S$ for later spectral types.
One additional contribution in the same sense can come from
chromospheric activity, as we describe in the next section.
Because all these three tendencies could not be properly
characterized, we dealt with the observed trend in the derived 
metallicities by relating our results to metallicities
determined using the Str\"{o}mgren photometry. This is 
described in Sections~\ref{determine}, \ref{Strom} 
and \ref{results}.

\section{Can photospheric and chromospheric activity
affect strengths of the lines?}
\label{appendix-C}

The magnesium triplet was observed to be moderately filled-in
by emission in spectra of solar flares by \citet{Acampa1982}. Similar
filling-in of the absorption lines was detected in the most 
chromospherically active stars \citet{Waite2011}. An artificial
reduction in the depth of the magnesium lines would lead to reduced
values of $S$ and thus 
derivation of too low metallicity. As was discussed in 
Section~\ref{spectra}, the magnesium-triplet lines contribute
very little ($<10$\%) to the overall line strength $S$ in the early
F-type spectra which are dominated by iron lines.  
But the triplet contribution grows for later spectral 
types so that -- if its lines are partly filled-in -- the metallicity may 
be biased. We stress that we do not assume that activity increases
for later spectral types; we only point out that the magnesium
triplet lines contribute more to the measured strengths $\log S$. 
The filling-in by emission cannot be determined 
from the EW binary spectra because there is no way of 
separating the magnesium 
lines from the multitude of rotationally blended iron lines in the spectral
window.

While a large fraction of numerous reports of light curve asymmetries 
in EW binaries may manifest the contact model deficiencies
\citep{prib2011}, some support for small, dark spots comes from spectroscopic 
BF profile observations \citep{Barnes2004,PR2008}.
It is important to stress
that as long as the spots are dark, they would not affect
our metallicity analysis at all. This is because we use rectified spectra 
so that the line depths are relative to the total flux
from the star, irrespectively of any flux reduction due to spots. 
The spots would influence our results only if they modified 
the observed colors (leading to wrong assumed $T_{\rm eff}$)
or contributed their own spectral signatures. While
the former effect cannot be excluded (but would require a separate
investigation of binaries with the definite presence of spots),  
the spectral signatures of spots are not expected to be detectable in 
low $S/N$ spectra of the magnesium triplet region. 
We note that detection of spot spectra in active stars has so 
far been attempted using special techniques in application to
molecular bands in the infra-red region \citep{IRspectra}.

\begin{deluxetable}{cll}

\tabletypesize{\scriptsize}

\rotate

\tablewidth{0pt}
\tablenum{1}
\tablecaption{The DDO spectroscopic survey. \label{tab_papers}}
\tablehead{
  \colhead{D1}         &
  \colhead{Binary systems} &
  \colhead{Reference}     
}
\startdata
0a  & \underline{AH~Vir}                  & \cite{LR1993}       \\
0b  & {\it W~Crv}                         & \cite{RL2000}       \\
1  & GZ~And, V417~Aql, [LS~Del], [EF~Dra],
     V829~Her, FG~Hya, [AP~Leo], [UV~Lyn],    
     BB~Peg, AQ~Psc                      & \cite{ddo1}          \\
2  & AH~Aur, CK~Boo, DK~Cyg, {\it SV~Equ}, 
     V842~Her, UZ~Leo, XZ~Leo, V839~Oph,
     GR~Vir, NN~Vir                      & \cite{ddo2}          \\
3  & {\it CN~And}, HV~Aqr, AO~Cam, YY~CrB,
     FU~Dra\tablenotemark{a}, {\it RZ~Dra}, UX~Eri, RT~LMi, 
     V753~Mon, OU Ser                    & \cite{ddo3}          \\
4  & \underline{44~Boo}, FI~Boo, V2150~Cyg, \underline{V899~Her},
     EX~Leo, \underline{VZ~Lib}, \underline{\it SW~Lyn}, V2377~Oph, 
     {\it DV~Psc}, \underline{HT~Vir}    & \cite{ddo4}          \\
5  & V376~And, EL~Aqr, EF~Boo, DN~Cam,
     FN~Cam, V776~Cas, SX~Crv, V351~Peg,
     EQ~Tau, {\it KZ~Vir}                & \cite{ddo5}          \\
6  & {\it SV~Cam}, EE~Cet, \underline{KR~Com}, V401~Cyg, 
     GM~Dra, V972~Her, ET~Leo, {\it FS~Leo},
     \underline{V2388~Oph}, \underline{II~UMa} & \cite{ddo6}    \\ 
8  & \underline{V410~Aur}, V523~Cas, QW~Gem, V921~Her,
     V2357~Oph, V1130~Tau, HN~UMa, HX~UMa,
     VY~Sex, DZ~Psc                      & \cite{ddo8}          \\
9  & AB~And, V402~Aur, V445~Cep, V2082~Cyg,
     BX~Dra, V918~Her, V502~Oph, V1363~Ori,
     \underline{KP~Peg}, V335~Peg        & \cite{ddo9}          \\
10 & {\it V395~And}, {\it HS~Aqr}, V449~Aur, FP~Boo,
     SW~Lac, {\it KS~Peg}, {\it IW~Per}, \underline{V592~Per},
     \underline{TU~UMi}, {\it FO~Vir}    & \cite{ddo10}         \\
11 & DU~Boo, \underline{\it ET~Boo}, TX~Cnc, V1073~Cyg,
     {\it HL~Dra}, AK~Her, \underline{VW~LMi}, V566~Oph,
     \underline{TV~UMi}, AG~Vir          & \cite{ddo11}         \\
12 & OO~Aql, CC~Com, V345~Gem, \underline{XY~Leo},
     AM~Leo, {\it V1010~Oph}, V2612~Oph, XX~Sex,
     W~UMa, {\it XY~UMa}                 & \cite{ddo12}         \\
13 & {\it EG~Cep}, V1191~Cyg, V1003~Her, \underline{BD+7~3142},
     V357~Peg, V407~Peg, V1123~Tau, V1128~Tau,
     HH~UMa, \underline{PY~Vir}          & \cite{ddo13}         \\
14 & \underline{TZ~Boo}, {\it VW~Boo}, \underline{EL~Boo}, {\it VZ~CVn}, 
     {\it GK~Cep}, RW~Com, \underline{V2610~Oph}, \underline{V1387~Ori}, 
     AU~Ser, \underline{FT~UMa}                      & \cite{ddo14}         \\
15 & QX And, DY~Cet, \underline{\it MR~Del}, HI~Dra, 
     \underline{\it DD~Mon}, V868~Mon, \underline{ER~Ori}, 
     \underline{Y~Sex}                   & \cite{ddo15}  \\    
   & {\it TT~Cet}, \underline{AA~Cet}, CW~Lyn, \underline{V563~Lyr},
     {\it CW~Sge}, \underline{LV~Vir}, {\it MW~Vir},
     {\it GO~Cyg}, {\it V857~Her},{\it V752~Mon},
     \underline{\it V353~Peg}              & ~~~~incomplete orbits   \\
16 & [V404~Peg], V407~Peg\tablenotemark{b}, [HH~Boo]     & \cite{ML2004}       \\
17 & {\it V471~Tau}                      & \cite{KK2007}       \\
18 & AW~UMa                              & \cite{PR2008}       \\
19 & \underline{GSC1387-475}             & \cite{RP2008}       \\
20 & {\it HD~73709}, {\it GSC~0814-0323} & \cite{MOST}         \\
21 & {\it BX~And}, {\it DO~Cas}, {\it BV~Eri}, 
     {\it VV~Cet}, {\it WZ~Cyg} & \cite{Siwak2010} 
\enddata
\tablecomments{
The papers of the DDO series have numbers D1 
between 1 and 15; additional papers included in the program have 
been assigned here arbitrary numbers outside the range 1 -- 15.
The paper DDO--7 \citep{ddo7} is not listed here because it
presented the method and a preliminary discussions at a mid-point of the
DDO survey; it contained no data. \\
Binaries excluded from the present study are marked in the table 
in three ways: In {\it italics\/} when non-contact,
\underline{underlined} when with a bright companion, $L_3/L_{12} > 0.05$, 
and in [square brackets] when the data were too poor to be used 
or the original spectra were lost.
}
\tablenotetext{a}{FU~Dra: Has also a variable-star name VX~Dra.}
\tablenotetext{b}{V407~Peg: More extensive data were published in DDO-13.}

\end{deluxetable}

\vfill
\newpage

\begin{deluxetable}{llr}

\tabletypesize{\scriptsize}

\tablewidth{0pt}

\tablenum{2}
\tablecaption{Spectral lines in the window 5080 -- 5285 \AA\   
\label{tab_lines}
}

\tablehead{
  \colhead{$\lambda$ \AA }         &
  \colhead{El.ion}            &
  \colhead{$W$~(m\AA)}     
}
\startdata
5110.413	  &  	Fe  I	  &  	80	 \\
5133.681	  &  	Fe  I	  &  	85	 \\
5139.251	  &  	Fe  I	  &  	89	 \\
5139.462	  &  	Fe  I	  &  	95	 \\
5142.927	  &  	Fe  I	  &  	98	 \\
5162.292	  &  	Fe  I	  &  	82	 \\
5165.407	  &  	Fe  I	  &  	80	 \\
5167.321	  &  	Mg  I	  &    135	 \\
5167.487	  &  	Fe  I	  &    112	 \\
5169.033	  &  	Fe  II	  &    853	 \\
5171.595	  &  	Fe  I	  &     99	 \\
5172.684	  &  	Mg  I	  &    270	 \\
5183.604	  &  	Mg  I	  &    356	 \\
5188.680	  &  	Ti  II	  &  	96	 \\
5191.455	  &  	Fe  I	  &  	92	 \\
5192.343	  &  	Fe  I	  &  	96	 \\
5194.941	  &  	Fe  I	  &  	89	 \\
5195.468	  &  	Fe  I	  &  	81	 \\
5197.577	  &  	Fe  II	  &    100	 \\
5202.335	  &  	Fe  I	  &  	80	 \\
5204.506	  &  	Cr  I	  &    100	 \\
5206.038	  &  	Cr  I	  &    105	 \\
5208.419	  &  	Cr  I	  &    109	 \\
5216.274	  &  	Fe  I	  &  	86	 \\
5226.543	  &  	Ti  II	  &  	94	 \\
5226.862	  &  	Fe  I	  &  	92	 \\
5227.150	  &  	Fe  I	  &  	93	 \\
5227.189	  &  	Fe  I	  &    118	 \\
5232.939	  &  	Fe  I	  &    106	 \\
5234.625	  &  	Fe  II	  &    103	 \\
5239.813	  &  	Sc  II	  &    112	 \\
5258.842	  &  	Si  I	  &  	84	 \\
5265.556	  &  	Ca  I	  &  	80	 \\
5266.555	  &  	Fe  I	  &  	98	 \\
5269.537	  &  	Fe  I	  &    126	 \\
5270.270	  &  	Ca  I	  &  	89	 \\
5270.357	  &  	Fe  I	  &    105	 \\
5276.002	  &  	Fe  II	  &    107	 \\
5281.700	  &  	Fe  I	  &  	83	 \\
5283.621	  &  	Fe  I	  &  	90	 
\enddata

\tablecomments{The equivalent widths $W$ of the lines with $W > 80$ m\AA\
for a model atmosphere approximating the template, HD~128167, with 
 $T_{\rm eff} = 7,000$~K and $\log g = 4.5$.}

\end{deluxetable}



\begin{deluxetable}{clcrccc}

\tabletypesize{\scriptsize}

\tablewidth{0pt}

\tablecaption{Basic assumed parameters for the analyzed EW binaries
\label{tab_main1}
}

\tablenum{3}

\tablehead{
\colhead{D1:D2} & \colhead{Name} & \colhead{Period} & 
          \colhead{$V_{\rm max}$} & \colhead{$B-V$} & 
          \colhead{$E(B-V)$} & \colhead{$d$}  \\ 
\colhead{}      & \colhead{}     & \colhead{(day)}  & \colhead{}  &
\colhead{}      & \colhead{}         & \colhead{(pc)} 
} 

\startdata
1:1 & GZ And   & 0.3050 & 10.85 &  [$1.12\pm0.10$] & 0.03 & $160 \pm 32 $ \\
1:2 & V417 Aql & 0.3703 & 10.62 &   $0.64\pm0.01$  & 0.05 & $200 \pm 40 $ \\
1:5 & V829 Her & 0.3582 & 10.39 &   $0.65\pm0.04$  & 0.01 & $210 \pm 42 $ \\
1:6 & FG Hya   & 0.3278 & 10.00 &   $0.62\pm0.01$  & 0.01 & $200 \pm 40 $ \\
1:9 & BB Peg   & 0.3615 & 10.97 &  [$0.48\pm0.01$] & 0.00 & $290 \pm 58 $ \\
1:10 & AQ Psc  & 0.4756 & 8.66  &  [$0.47\pm0.01$] & 0.04 & $150 \pm 23 $ \\
2:1 & AH Aur   & 0.4943 & 10.18 &   $0.64\pm0.01$  & 0.03 & $110 \pm 22 $ \\
2:2 & CK Boo   & 0.3552 & 9.09  &   $0.52\pm0.01$  & 0.02 & $120 \pm 24 $ \\
2:3 & DK Cyg   & 0.4707 & 10.47 &   $0.38\pm0.01$  & 0.03 & $120 \pm 24 $ \\
2:5 & V842 Her & 0.4190 & 9.98  &   $0.67\pm0.04$  & 0.02 & $240 \pm 48 $
\enddata
\tablecomments{
The publication identification number D1:D2 uses the paper number D1
(see Table~\ref{tab_papers}) and the successive number D2 within the
paper. 
The photometric data $V_{\rm max}$, $B-V$ and $E(B-V)$ and 
the distances ($d$) 
have been estimated as described in the text. 
The $B-V$ colors taken from the Tycho-2 Catalog \citep{esa1997}
are given in square brackets. \\
The full table is published electronically.}

\end{deluxetable}

\vfill
\newpage


\begin{deluxetable}{crrrrrrc}

\tabletypesize{\scriptsize}

\tablewidth{0pt}
\tablenum{4}
\tablecaption{Model atmosphere values of $\log S$ \label{tab_models}}
\tablehead{
\colhead{$B-V$}     &
\colhead{$-2.0$}    &
\colhead{$-1.5$}    &
\colhead{$-1.0$}    &
\colhead{$-0.5$}    &
\colhead{$0.0$}     &
\colhead{$+0.5$}    &
\colhead{$\Delta_g \log S$}
}
\startdata
0.00  & $-1.286$ &  $-1.038$ & $-0.820$  & $-0.620$  & $-0.428$  & $-0.232$  & $-0.028$ \\
0.10  & $-1.037$ &  $-0.807$ & $-0.602$  & $-0.418$  & $-0.252$  & $-0.099$  & $-0.019$ \\
0.20  & $-0.805$ &  $-0.591$ & $-0.402$  & $-0.237$  & $-0.095$  &  0.023  & $-0.010$ \\
0.30  & $-0.591$ &  $-0.393$ & $-0.221$  & $-0.075$  &  0.043  &  0.132  & $-0.010$ \\
0.40  & $-0.396$ &  $-0.212$ & $-0.058$  &  0.067  &  0.163  &  0.228  & $-0.014$ \\
0.50  & $-0.220$ &  $-0.050$ & 0.086  &  0.190  &  0.265  &  0.313  & $-0.024$ \\
0.60  & $-0.064$ &   0.092 &  0.210  &  0.294  &  0.350  &  0.384  & $-0.031$ \\
0.70  &  0.072 &   0.214 &  0.313  &  0.379  &  0.419  &  0.443  & $-0.036$ \\
0.80  &  0.188 &   0.314 &  0.397  &  0.446  &  0.473  &  0.489  & $-0.033$ \\
0.90  &  0.282 &   0.392 &  0.459  &  0.495  &  0.512  &  0.522  & $-0.026$ \\
1.00  &  0.355 &   0.447 &  0.500  &  0.526  &  0.536  &  0.542  & $-0.015$ \\
1.10  &  0.407 &   0.478 &  0.520  &  0.539  &  0.546  &  0.549  & $-0.006$ \\
1.20  &  0.436 &   0.484 &  0.520  &  0.539  &  0.546  &  0.549  & $-0.002$ 
\enddata
\tablecomments{The table lists the model values of
$\log S$ for $\log g = 4.5$ and
for the values of $[M/H]$ as given in the column headers labeled 
$-2.0$ to $+0.5$. The median
differences in $\log S$ (for all metallicities) 
between $\log S$ for $\log g = +3.5$ and $+4.5$,
scaled to $\Delta \log g = 0.3$, are given in the last column.
}
\end{deluxetable}

\vfill
\newpage


\begin{deluxetable}{clcrrcrcccc}

\tabletypesize{\footnotesize}

\tablewidth{0pt}


\tablecaption{Measured combined line strengths
and metallicities \label{tab_main2}}

\tablenum{5}

\tablehead{
\colhead{D1:D2}  & \colhead{Name} & \colhead{Gr}         & \colhead{m}       & 
\colhead{$\log S$} & \colhead{$\epsilon (\log S)$} & \colhead{$\delta \times 10^3$} & 
\colhead{$[M/H]$} & \colhead{$[M/H]_{\rm m}$}  & \colhead{$[M/H]_1$} & \colhead{$\epsilon [M/H]$}
} 

\startdata
1:1 & GZ And    & R & 26 & $0.394\pm0.004$ & 0.100 & $-0.95\pm0.29$  & [$-2.03$] & \nodata & [$-0.52$] & [$0.28$] \\
1:2 & V417 Aql  & F & 22 & $0.229\pm0.007$ & 0.076 & $-0.20\pm0.27$  & $-0.84$ & $-0.46$ & $-0.46$ & $0.19$ \\
1:5 & V829 Her  & R & 4 & $0.249\pm0.015$ & 0.082 & $-2.24\pm0.43$  & $-0.90$ & \nodata & $-0.43$ & $0.21$ \\
1:6 & FG Hya    & F & 30 & $0.257\pm0.005$ & 0.074 & $-2.68\pm0.17$  & $-0.67$ & $-0.14$ & $-0.26$ & $0.19$ \\
1:9 & BB Peg    & F & 28 & $0.298\pm0.006$ & 0.058 & $-0.13\pm0.27$  & [$0.51$] & \nodata & [$0.63$] & [$0.15$] \\
1:10 & AQ Psc    & F & 37 & $0.202\pm0.003$ & 0.056 & $-2.56\pm0.07$  & $0.18$ & \nodata & $0.18$ & $0.14$ \\
2:1 & AH Aur    & F & 32 & $0.296\pm0.004$ & 0.077 & $-2.55\pm0.16$  & $-0.41$ & $0.08$ & $0.00$ & $0.19$ \\
2:2 & CK Boo    & F & 28 & $0.179\pm0.004$ & 0.061 & $-2.04\pm0.14$  & $-0.49$ & \nodata & $-0.33$ & $0.15$ \\
2:3 & DK Cyg    & F & 27 & $0.016\pm0.005$ & 0.041 & $1.12\pm0.16$  & $-0.42$ & $-0.26$ & $-0.61$ & $0.10$ \\
2:5 & V842 Her  & R & 26 & $0.324\pm0.006$ & 0.083 & $-2.95\pm0.20$  & $-0.42$ & \nodata & $0.08$ & $0.22$
\enddata

\tablecomments{The column headings which were not explained for
Table~\ref{tab_main1}:
D1:D2 -- binary identification; 
Gr -- the group to which the binary belongs: 
B for blue ($(B-V)_0 \le 0.32$), F for F-type ($0.32 < (B-V)_0 \le 0.62$),
and R for red ($(B-V)_0 > 0.62$); 
m -- the number of phase points used for averaging of
independent orbital points;
$\log S$ -- the mean Broadening Function strength with its observational error 
($\epsilon_1$) evaluated from the phase scatter;
$\epsilon (\log S)$ -- the assumed, combined error of $\log S$ consisting of 
five components, as described in Appendix~\ref{errors}; 
$\delta$ -- the measured shift in the BF baseline with 
its error, both multiplied by $10^3$ for more convenient formatting.\\
The metallicities: the raw determination $[M/H]$, 
the metallicity derived from the Str\"omgren $m_1$ index $[M/H]_{\rm m}$, 
the trend-corrected $[M/H]_1$ 
and the assumed final error $\epsilon [M/H]$. Metallicities determined
through extrapolation beyond the model grid and thus entirely 
unreliable are given in square brackets (see the text).\\
The full table is published electronically.}

\end{deluxetable}

\vfill
\newpage


\begin{deluxetable}{clcccc}

\tabletypesize{\footnotesize}

\tablewidth{0pt}

\tablecaption{Str\"{o}mgren $uvby$ photometry literature data \label{tab_uvby}}

\tablenum{6}

\tablehead{
\colhead{D1:D2} & \colhead{Name} & \colhead{$b-y$} & 
\colhead{$m_1$} & \colhead{$c_1$} & \colhead{Source}  \\ 
\colhead{} & \colhead{} & \colhead{} & \colhead{} & \colhead{} & \colhead{} 
} 

\startdata
1:2 & V417~Aql & 0.421 & 0.146 & 0.360 & 1 \\
1:6 & FG~Hya & 0.385 & 0.182 & 0.333 & 1 \\
2:1 & AH~Aur & 0.401 & 0.195 & 0.423 & 1 \\
2:3 & DK~Cyg & 0.248 & 0.137 & 0.672 & 1 \\
2:6 & UZ~Leo & 0.242 & 0.143 & 0.685 & 1 \\
2:8 & V839~Oph & 0.399 & 0.206 & 0.376 & 1 \\
2:9 & GR~Vir & 0.370 & 0.157 & 0.352 & 2 \\
2:10 & NN~Vir & 0.248 & 0.165 & 0.579 & 2 \\
3:4 & YY~CrB & 0.368 & 0.172 & 0.341 & 3 \\
3:9 & V753~Mon & 0.213 & 0.152 & 0.704 & 2 \\
3:10 & OU~Ser & 0.405 & 0.152 & 0.293 & 3 \\
4:5 & EX~Leo & 0.331 & 0.153 & 0.395 & 2 \\
4:8 & V2377~Oph & 0.407 & 0.172 & 0.322 & 2 \\
5:4 & DN~Cam & 0.216 & 0.163 & 0.631 & 2 \\
5:8 & V351~Peg & 0.199 & 0.168 & 0.735 & 4 \\
6:6 & V972~Her & 0.295 & 0.152 & 0.469 & 2 \\
8:4 & V921~Her & 0.237 & 0.141 & 0.789 & 4 \\
8:6 & V1130~Tau & 0.276 & 0.124 & 0.474 & 2 \\
8:8 & HX~UMa & 0.309 & 0.158 & 0.366 & 5 \\
9:1 & AB~And & 0.510 & 0.373 & 0.320 & 1 \\
9:4 & V2082~Cyg & 0.224 & 0.166 & 0.698 & 2 \\
9:10 & V335~Peg & 0.312 & 0.163 & 0.434 & 2 \\
10:3 & V449~Aur & 0.086 & 0.163 & 1.019 & 6 \\
10:5 & SW~Lac & 0.471 & 0.284 & 0.268 & 1 \\
11:3 & TX~Cnc & 0.388 & 0.206 & 0.348 & 1 \\
11:4 & V1073~Cyg & 0.281 & 0.170 & 0.666 & 1 \\
11:6 & AK~Her & 0.342 & 0.176 & 0.397 & 1 \\
11:8 & V566~Oph & 0.305 & 0.142 & 0.445 & 2 \\
11:10 & AG~Vir & 0.156 & 0.166 & 0.824 & 1 \\
12:1 & OO~Aql & 0.464 & 0.261 & 0.291 & 1 \\
12:5 & AM~Leo & 0.356 & 0.177 & 0.341 & 1 \\
12:7 & V2612~Oph & 0.382 & 0.149 & 0.361 & 7 \\
12:9 & W~UMa & 0.406 & 0.203 & 0.292 & 1 \\
15:1 & QX~And & 0.325 & 0.156 & 0.421 & 8 \\
18:1 & AW~UMa & 0.246 & 0.142 & 0.602 & 9,10 \\
\enddata

\tablerefs{1. \citet{ruckal1981};
2. \citet{ols1983};
3. \citet{ols1993};
4. \citet{perr1982};
5. \citet{blaa1976};
6. \citet{jord1996};
7. \citet{schm1976};
8. \citet{craw1970};
9. \citet{hauck1998};
10. \citet{holm2009}}

\end{deluxetable}

\vfill
\newpage


\begin{deluxetable}{clrrrr}

\tabletypesize{\footnotesize}

\tablewidth{0pt}

\tablecaption{Kinematic input data \label{tab_kinem1}}

\tablenum{7}

\tablehead{ 
\colhead{D1:D2} & \colhead{Name} & \colhead{HIP} & \colhead{$V_0$} 
& \colhead{$\mu$RA} & \colhead{$\mu$Dec} \\ 
\colhead{} & \colhead{} & \colhead{} &
\colhead{(km~s$^{-1}$)} & \colhead{(mas/yr)} & \colhead{(mas/yr)} 
} 

\startdata
  1:2 & V417 Aql   &  96349 & $-14.20 \pm 1.40$ & $ -2.24 \pm 2.76$ & $-24.70 \pm 1.63$ \\
  1:6 & FG Hya     &  41437 & $ -5.70 \pm 1.20$ & $  5.06 \pm 1.95$ & $-63.72 \pm 1.56$ \\
  1:9 & BB Peg     & 110493 & $-28.90 \pm 1.40$ & $ 20.00 \pm 2.42$ & $-25.56 \pm 1.68$ \\
 1:10 & AQ Psc     &   6307 & $-12.90 \pm 0.40$ & $ -8.93 \pm 1.08$ & $ 14.25 \pm 0.73$ \\
  2:1 & AH Aur     &  30618 & $ 31.95 \pm 1.45$ & $ 15.23 \pm 2.72$ & $-12.50 \pm 1.64$ \\
  2:2 & CK Boo     &  71319 & $ 37.04 \pm 0.75$ & $ 71.05 \pm 1.05$ & $-87.41 \pm 1.08$ \\
  2:3 & DK Cyg     & 106574 & $ -5.53 \pm 1.48$ & $-28.27 \pm 0.91$ & $-19.24 \pm 1.14$ \\
  2:6 & UZ Leo     &  52249 & $ -8.18 \pm 1.12$ & $-21.35 \pm 1.25$ & $ -1.88 \pm 0.75$ \\
  2:7 & XZ Leo     &  49204 & $ -2.03 \pm 1.43$ & $-16.49 \pm 1.95$ & $  3.00 \pm 1.01$ \\
  2:8 & V839 Oph   &  88946 & $-63.81 \pm 1.09$ & $-34.53 \pm 1.68$ & $  2.82 \pm 1.54$
\enddata

\tablecomments{The column headings which were not explained in the
previous tables: 
HIP -- the Hipparcos mission number (HD numbers
and other names are given in the respective DDO papers); 
$V_0$ -- center-of-mass radial velocity from the DDO survey. \\
Proper motions $\mu$RA and $\mu$Dec 
have been taken from the new
reductions of the Hipparcos data by \citet{leeu2007}.\\
The full table is published electronically.}

\end{deluxetable}

\vfill
\newpage

\begin{deluxetable}{clccc}

\tabletypesize{\footnotesize}

\tablewidth{0pt}

\tablecaption{Space velocity components \label{tab_kinem2}}

\tablenum{8}

\tablehead{\colhead{D1:D2} & \colhead{Name} &
   \colhead{U} & \colhead{V} & \colhead{W} \\ 
           \colhead{} & \colhead{} &
   \colhead{(km~s$^{-1}$)} & \colhead{(km~s$^{-1}$)} & \colhead{(km~s$^{-1}$)} } 

\startdata
  1:1 &  GZ And     & [$ 14.0 \pm  1.9$] & [$ 21.1 \pm  1.9$] & [$  2.4 \pm  2.0$] \\ 
  1:2 &  V417 Aql   &  $ 12.2 \pm  4.4 $ &  $-12.8 \pm  4.6 $ &  $ -0.9 \pm  5.1 $ \\ 
  1:5 &  V829 Her   & [$  5.6 \pm  1.0$] & [$ -0.5 \pm  1.0$] & [$  0.8 \pm  1.1$] \\ 
  1:6 &  FG Hya     &  $ 41.9 \pm  6.6 $ &  $-31.2 \pm 10.0 $ &  $-18.0 \pm  5.6 $ \\ 
  1:9 &  BB Peg     &  $  0.7 \pm  6.1 $ &  $-34.8 \pm  6.1 $ &  $-14.7 \pm  8.9 $ \\ 
 1:10 &  AQ Psc     &  $ 14.6 \pm  1.4 $ &  $ 19.0 \pm  2.1 $ &  $ 22.0 \pm  1.2 $ \\ 
  2:1 &  AH Aur     &  $-21.7 \pm  2.9 $ &  $  1.2 \pm  2.7 $ &  $ 14.7 \pm  2.8 $ \\ 
  2:2 &  CK Boo     &  $ 81.8 \pm 11.0 $ &  $  2.9 \pm  2.5 $ &  $  6.6 \pm  6.5 $ \\ 
  2:3 &  DK Cyg     &  $ 27.0 \pm  4.0 $ &  $  6.5 \pm  2.9 $ &  $ 10.8 \pm  1.5 $ \\ 
  2:5 &  V842 Her   & [$-25.3 \pm  3.4$] & [$-33.9 \pm  2.0$] & [$-23.7 \pm  2.3$]
\enddata

\tablecomments{The table lists new space velocity 
determinations except for velocities from \citet{bilir2005}
which are given in square brackets (after the LSR correction, as described
in the text). The remaining columns have column headings as in other tables. \\
The full table is published electronically.}

\end{deluxetable}

\vfill
\newpage

\end{document}